\documentclass[twocolumn,showpacs,preprintnumbers,amsmath,amssymb]{revtex4}
\usepackage{graphicx} 
\usepackage{dcolumn}  
\usepackage{bm}       
\begin{document} 
\title{Self-energy and Fermi surface of the 2-dimensional Hubbard model}
\author{R. Eder$^{1,2}$,K. Seki$^1$, and Y. Ohta$^1$}
\affiliation{$^2$Department of Phsics, Chiba University, Chiba 263-8522, Japan\\
$^2$Karlsruhe Institut of Technology,
Institut f\"ur Festk\"orperphysik, 76021 Karlsruhe, Germany}
\date{\today}
\begin{abstract}
We present an exact diagonalization study of the self-energy of the 
two-dimensional Hubbard model. To increase the range of available cluster 
sizes we use a corrected t-J model to compute approximate Greens functions 
for the Hubbard model.  This allows to obtain spectra for clusters with
18 and 20 sites. The self-energy has several `bands' of poles with 
strong dispersion and extended incoherent continua with k-dependent
intensity. We fit the  self-energy by a minimal model and use 
this to extrapolate the cluster results to the infinite lattice. The resulting 
Fermi surface shows a  transition from hole pockets in the underdoped regime 
to a large Fermi surface in the overdoped regime. We demonstrate that
hole pockets can be completely consistent with the Luttinger
theorem. Introduction of next-nearest neighbor hopping changes the self-energy
stronlgy and the spectral function with nonvanishing next-nearest-neighbor 
hopping in the underdoped region is in good agreement with angle resolved 
photoelectron spectroscopy. 
\end{abstract} 
\maketitle
\section{Introduction}
Experiments on cuprate superconductors have shown a nontrivial
evolution of their Fermi surface with hole doping $\delta$.
In the overdoped compound Tl$_2$Ba$_2$CuO$_{6+\delta}$
magnetoresistance measurements\cite{Hussey}, angle-resolved photoemission
spectroscopy (ARPES)\cite{Plate} and quantum 
oscillation experiments\cite{Vignolle} show a fairly conventional
Fermi surface which is consistent with LDA band structure
calculations that take the Cu3d electrons as itinerant
and which covers a fraction of the Brillouin
zone of $\approx (1-\delta)/2$.
In the underdoped compounds the situation is more involved. ARPES shows
`Fermi arcs'\cite{Damascelli} which however are probably just the intense
part of hole pockets centered near $(\frac{\pi}{2},\frac{\pi}{2})$.
This is plausible because
the sharp drop of the ARPES weight of the  quasiparticle band
upon crossing the noninteracting Fermi surface which must be invoked
to reconcile the `Fermi arcs' with the hole pocket scenario 
is actually well established in insulating cuprates 
such as Sr$_2$Cu$_2$O$_2$Cl$_2$\cite{Wells} and
Ca$_2$CuO$_2$Cl$_2$\cite{Ronning} where this phenomenon
has been termed the 'remnant Fermi surface'.
Moreover  both the Drude weight
in La$_{2-x}$Sr$_x$CuO$_4$\cite{Uchida,Padilla}
and YBa$_2$Cu$_3$O$_y$\cite{Padilla} as well as the inverse
low temperature Hall constant in 
La$_{2-x}$Sr$_x$CuO$_4$\cite{Ong,Takagi,Ando,Padilla} and
YBa$_2$Cu$_3$O$_y$\cite{Padilla}
scale with $\delta$ and the inferred band mass is 
constant troughout the underdoped regime and in fact even the 
antiferromagnetic phase\cite{Padilla}. 
This would be a exactly the behaviour expected for
hole pockets. On the other hand, 
for $\delta \ge 0.15$ the Hall constant in La$_{2-x}$Sr$_x$CuO$_4$
changes rapidly, which suggests a change from hole pockets to a large
Fermi surface\cite{Ong}.
Quantum-oscillation experiments on underdoped
YBa$_2$Cu$_3$O$_{6.5}$\cite{Doiron,Sebastian_1,Jaudet,Audouard}
and YBa$_2$Cu$_4$O$_8$\cite{Yelland,Bangura}
show that the Fermi surface has a cross section that is comparable 
to $\delta/2$ rather than $(1-\delta)/2$. Thereby the mere
validity of the Fermi liquid description as evidenced by the 
quantum oscillations is clear evidence against the
notion of `Fermi arcs': the defining property of a Fermi liquid is the
one-to-one correspondence of its low-lying states to those of
a fictitious system of weakly interacting Fermionic quasiparticles
and the Fermi surface of these quasiparticles
is a constant energy contour of their dispersion
and therefore necessarily a closed curve in ${\bf k}$-space.
On the other hand the quantum oscillations cannot be viewed
as evidence for hole pockets either in that
both the Hall constant\cite{LeBoeuf} and thermopower\cite{Chang} 
have a sign 
that would indicate electron pockets in the normal state induced by the
high magnetic fields used in the quantum oscillation experiments.
Thereby both, the Hall constant and the thermopower, show a strong
temperature dependence and in fact a sign change
as a function of temperature. At the same
time neutron scattering experiments on detwinned
YBa$_2$Cu$_3$O$_{6.6}$ in the superconducting state
show anisotropy in the spin
excitations spectrum below $30\;meV$ and at low temperatures\cite{Hinkov}.
This indicates a rather complicated reconstruction to take place,
possibly to a `nematic' state with inequivalent
$x$- and $y$-direction in the CuO$_2$ plane. Such a
nematicity which is also apparent in scanning tunneling 
microscopy\cite{Lawler}
must modify the Fermi surface in some way which may explain the
unexpected sign.
All in all the data may be interpreted as showing a change of the Fermi surface
volume at around optimal doping from a small Fermi surface with
a volume $\propto \delta/2$ to a large one with volume 
$\propto (1-\delta)/2$.\\
Exact diagonalization studies of the t-J model
have shown that the Fermi surface at hole dopings
$\le 15\%$ takes the form of hole pockets\cite{poc1,poc2,poc3},
that the quasiparticles have the character of strongly renormalized
spin polarons throughout this doping range\cite{r1,r2,r3} and that the low 
energy spectrum at these doping levels can be described as a Fermi liquid of 
spin $1/2$ quasiparticles corresponding to the doped holes\cite{lan}.
A comparison of the dynamical spin and density correlation function
at low\cite{den1,den} ($\delta < 15\%$)
and intermediate  and high ($\delta=30-50 \%$) hole doping moreover
indicates\cite{intermediate} that around optimal doping a phase 
transition takes place.
In the underdoped regime spin and density correlation function differ strongly, 
with magnon-like spin excitations and extended incoherent continua in the
density correlation function\cite{den1,den} which can be explained 
quantitatively by a calculation in the spin-polaron 
formalism\cite{beckervoijta}. At higher doping, spin and density correlation 
function become more and more similar and both approach the self-convolution 
of the single-particle Green's function, whereby deviations from the 
self-convolution form can be explained as particle-hole excitations across a 
free electron-like Fermi surface\cite{intermediate}.
We thus expect a transition between a low-doping phase with
a hole-pocket Fermi surface and quasiparticles which resemble
the spin polarons realized at half-filling and a high doping phase
with a free electron-like large Fermi surface.
Here we want to further elucidate the issue of the Fermi surface and
the possible transition between the large and small Fermi surfaces.
To that end we study the electronic self-energy $\Sigma({\bf k},\omega)$
of the 2D Hubbard model by exact diagonalization.
\section{Model and Method of calculation}
We study the Hubbard model on a two dimensional
square lattice, defined by the Hamiltonian
\begin{equation}
H=-t\sum_{\langle i,j \rangle}\sum_\sigma\;\left( c_{i,\sigma}^\dagger
c_{j,\sigma}^{} + H.c.\right) + U \sum_i n_{i,\uparrow}n_{i,\downarrow}.
\end{equation}
Here $c_{i,\sigma}^\dagger$ creates an electron with $z$-spin $\sigma$
in the orbital at lattice site $i$ and $\langle i,j \rangle$ denotes a summation
over all nearest neighbor pairs.
We set $t=1$ and unless otherwise stated $U/t=10$.
The self-energy $\Sigma({\bf k},\omega)$ is defined by the Dyson equation
\begin{equation}
G^{-1}({\bf k},\omega) = \omega  - \epsilon_{\bf k} 
- \Sigma({\bf k},\omega)
\label{self}
\end{equation} 
where $\epsilon_{\bf k}= -2t( \cos(k_x)+\cos(k_y))$ is the free dispersion
and
\begin{eqnarray}
G({\bf k},\omega) &=& \langle\Psi_0^{(N)}|c_{{\bf k},\sigma}^\dagger
\frac{1}{\omega - E_0^{(N)} + H}c_{{\bf k},\sigma}^{} + \nonumber \\
&&\;\;\;\;\;\;\;
c_{{\bf k},\sigma}^{}\frac{1}{\omega - H + E_0^{(N)}}c_{{\bf k},\sigma}^\dagger|\Psi_0^{(N)}\rangle
\label{green}
\end{eqnarray}
is the single particle Green's function at zero temperature\cite{Luttinger}. 
Here $|\Psi_0^{(N)}\rangle$ and
$E_0^{(N)}$ denote the ground state wave function
and energy with $N$ electrons. In the present study the Green's function
for finite clusters is evaluated numerically
by means of the Lanczos algorithm\cite{dagoreview}.\\
Luttinger\cite{Luttinger} has derived the following spectral representation
of the self-energy:
\begin{equation}
\Sigma({\bf k},\omega)= g_{\bf k} + 
\sum_{\nu} \frac{\sigma_{{\bf k},\nu}}{\omega - \zeta_{{\bf k},\nu}}.
\label{spectral}
\end{equation}
In other words $\Sigma({\bf k},\omega)$ is the sum of a real
constant $g_{\bf k}$ (which is equal to the
Hartree-Fock potential, see Appendix A) and a sum of poles on the real
axis. In the thermodynamical limit there may be both isolated
poles and  continua of poles $\zeta_{{\bf k},\nu}$.
In a finite system, however, the poles $\zeta_{{\bf k},\nu}$
in principle always are discrete.
Since the real part of
$\Sigma({\bf k},\omega)$ assumes any value in $[-\infty:\infty]$
in between two successive poles, $\zeta_{{\bf k},\nu}$ and
$\zeta_{{\bf k},\nu+1}$ it follows that the equation
\begin{equation}
\Re\;G^{-1}({\bf k},\omega)=0
\label{qpeq}
\end{equation}
has exactly one solution in the interval
$[\zeta_{{\bf k},\nu},\zeta_{{\bf k},\nu+1}]$.
If there is an energy interval with
zero spectral weight - i.e. a gap - in the single particle spectral
function it follows that there must be precisely
one pole of the self-energy within in this gap. For example,
the Hubbard-I approximation\cite{Hubbard}
 for a nonmagnetic ground state corresponds to 
\begin{equation}
\Sigma({\bf k},\omega) = nU + \frac{n(1-n)U^2}{\omega-(1-n)U}
\end{equation}
where $n$ is the density of electrons/spin.
This is a single ${\bf k}$-independent pole of strength $\propto U^2$ at 
approximately the center of the Hubbard gap. 
In the neighborhood of a pole of $\Sigma({\bf k},\omega)$
the real part of the self-energy takes the form
\[
\Sigma_r(\omega) + \frac{\sigma_{{\bf k},\nu}}{\omega - \zeta_{{\bf k},\nu}}.
\]
on the real axis, where $\Sigma_r(\omega)$ is slowly varying.
If the residuum $\sigma_{{\bf k},\nu}$ is large, the real part
is large as well and no solution of
$\omega  - \epsilon_{\bf k}= \Re\;\Sigma({\bf k},\omega)$
exists close to the pole. An isolated pole with large residuum thus
'pushes open' a gap of the spectral density in its neighborhood.
On the other hand, if $\sigma_{{\bf k},\nu}$ is small 
the real part will deviate from $\Sigma_r(\omega)$
only in the immediate neighborhood of $\zeta_{{\bf k},\nu}$.
This implies that the corresponding solution
of $\omega  - \epsilon_{\bf k}= \Re\;\Sigma({\bf k},\omega)$
is pinned near $\zeta_{{\bf k},\nu}$. Moreover, close to
$\zeta_{{\bf k},\nu}$ the slope of the real part of the self-energy
is large and negative, so that the spectral weight
$(1-\partial \Sigma({\bf k},\omega)/\partial \omega)^{-1}$ is small.
This rule will be seen frequently in the numerical spectra: an isolated pole
with large residuum opens a gap in the single-particle
spectral function around itself,
a pole with small residuum has a pole of the single-particle Green's
function with small weight in its immediate neighborhood.
Finally, we note that a `band' of poles of the self energy,
i.e. $\zeta_{{\bf k},\nu}$ with $\nu$ fixed, can never 
be crossed by a band of poles of the Green's function. Therefore,
bands of isolated poles of $\Sigma({\bf k},\omega)$
define surfaces in the three-dimensional
$(k_x,k_y,\omega)$-space which cannot be crossed by quasiparticle
bands in the Green's function. The only exception would be
a zero of the residuum, $\sigma_{{\bf k},\nu}$.\\
As already mentioned we study the self-energy  by computing
the Green's function of finite clusters by means of the Lanczos algorithm.
Thereby we encounter a technical problem concerning
the dimension of the Hilbert space. In a $4\times 4$ cluster
the dimension of the Hilbert space at half-filling (i.e. with
$8$ electrons of either spin direction in the cluster) is
$65636900$, in the half-filled 18-site cluster it is already $2363904400$.
Such large Hilbert space dimensions make numerical calculations
very difficult.
By contrast, the dimension of the Hilbert space of the Heisenberg model
- which is equivalent to the Hubbard model for large $U/t$ -
in the $4\times 4$ cluster is only $12870$,
in the $18$-site cluster it is $48620$.
The Heisenberg model - and in the doped case the t-J model - thus are
much easier to study numerically and in fact the
largest cluster for which exact diagonalization
studies for the Hubbard model have been performed\cite{ortolani,leung}
is $4\times 4$, whereas larger clusters are possible
for the t-J model.
For a study of the self energy, however, the t-J model cannot be used due 
to its `projected' nature
which for example implies the absence of the upper Hubbard band.\\
On the other hand, various authors have derived effective Hamiltonians
which operate in the projected Hilbert space of the t-J model but
reproduce physical quantities of the Hubbard model to order
$t/U$\cite{HarrisLange,Chao,MacDo,Eskesetal}.
This is achieved by performing a canonical transformation
which eliminates the part of the hopping term which creates/annihilates
double occupancies.
The crucial point thereby is that not only the Hamiltonian itself,
but all operators whose expectation values or correlation functions
are to be calculated, have to be subject to this canonical transformation
as well, which usually leads to correction terms of order $t/U$ in all
operators\cite{Eskesetal,EskesOles,EskesEder}.
If this is done consistently, however, very accurate approximate
spectra for the Hubbard model can be
calculated  using `t-J-sized' Hilbert spaces which allows to
treat more clusters and thus obtain additional 
information\cite{EskesEder}.
So far this procedure has been performed only for the lower
Hubbard band, because the study of the upper Hubbard band
in the doped case requires a considerable number
of additional terms in the Hamiltonian
which describe the interaction between
the doped holes and the double occupancy created in the inverse 
photoemission process\cite{Eskesetal}. For the present study, however, the
complete Hamiltonian as given by Eskes {\em et al.}\cite{Eskesetal} 
has been implemented as computer code. A brief outline of the
procedure and expressions for the corrected photoemission and
inverse photoemission operators are given in Appendix B. 
This procedure allows to calculate approximate
Green's function for the Hubbard model over 
the entire doping range and on all clusters for which t-J model calculations
are possible.
For all these systems we evaluated the single particle 
Greens function (\ref{green})
by the Lanczos method and obtained the self-energy from (\ref{self}).
\begin{figure}
\includegraphics[width=\columnwidth]{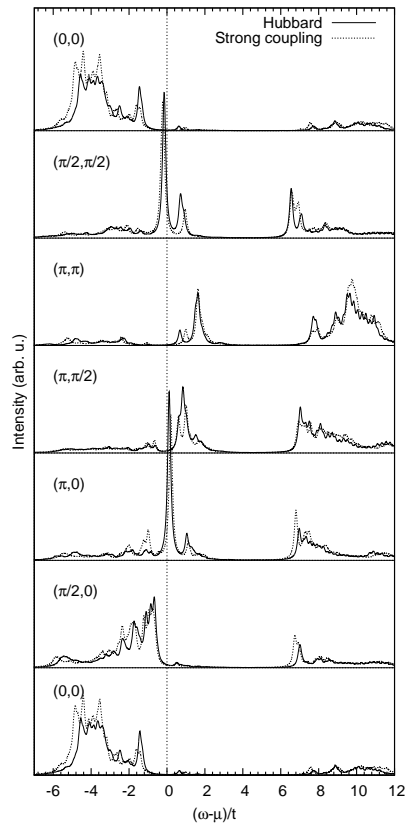}
\caption{\label{fig1}  
Spectral density $A({\bf k},\omega)$ for a $4\times 4$ cluster
with 2 holes
computed with the true Hubbard model and with the strong coupling model.
The ratio $U/t=10$, $\eta=0.1$.}
\end{figure}
To illustrate the accuracy that can be expected
Fig. \ref{fig1} compares the single particle
spectral function 
\begin{equation}
A({\bf k},\omega)= \frac{1}{\pi}\;\Im \;G({\bf k},\omega-i\eta)
\end{equation}
for the true Hubbard model and the strong coupling
Hamiltonian.
While there are clearly some small differences the strong coupling
model reproduces the spectral function of the Hubbard model
quite well. The deviations between the spectra calculated with the
strong coupling model and the true Hubbard model
are of order $t^3/U^2$ for energies and
$t^2/U^2$ for the weights. This property in fact
can be used to check the correctness of the strong-coupling-code
be comparing energies and weights of peaks at diferent $U/t$.
The agreement thus improves rapidly
with decreasing $t/U$ and already for $U/t=20$ the spectra become
essentially indistinguishable. It is therefore a useful check whether
certain features of the spectra are robust with decreasing $t/U$.\\
Lastly we mention that for the sake of analysis and extrapolation
to the infinite lattice the energies $\zeta_{{\bf k},\nu}$ and
residua $\sigma_{{\bf k},\nu}$ of some poles will frequently be expanded
in terms of tight-binding harmonics of ${\bf k}$ e.g.
\begin{eqnarray}
 \zeta_{{\bf k},\nu}&=& \sum_{j=0}^3 \zeta_{j,\nu} \gamma_{j}({\bf k})
\nonumber \\
\gamma_{0}({\bf k}) &=& 1
\nonumber \\
\gamma_{1}({\bf k}) &=& 2\cos(k_x)+2\cos(k_y)
\nonumber \\
\gamma_{2}({\bf k}) &=& 4\cos(k_x)\cos(k_y)
\nonumber \\
\gamma_{3}({\bf k}) &=& 2\cos(2k_x)+2\cos(2k_y)
\label{fit}
\end{eqnarray}
and similarly for the residua with coefficients $\sigma_{j,\nu}$ .
\section{Results for the Green's function and Self-energy}
\begin{figure}
\includegraphics[width=\columnwidth]{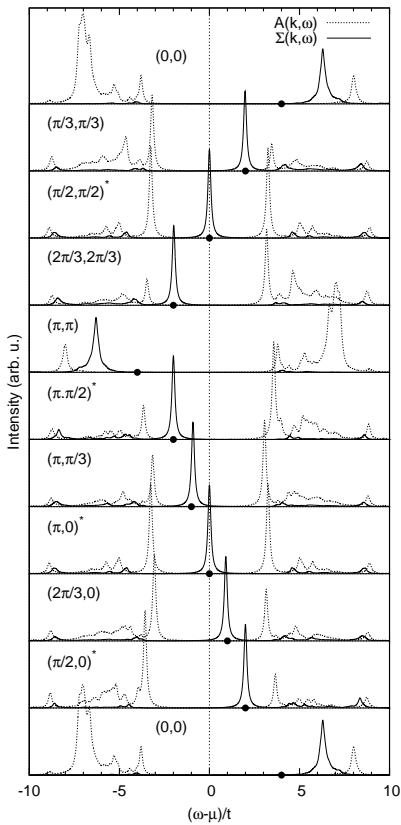}
\caption{\label{fig2}  
Spectral density $A({\bf k},\omega)$ and imaginary part of
the self-energy
$\Sigma({\bf k},\omega)$ at half-filling. The figure combines spectra
from the $16$- and $18$-site clusters, momenta from the
$16$-site cluster are marked by asterisks. The spectra are computed
with $\eta=0.1$, $\Sigma({\bf k},\omega)$ is multiplied by $1/50$. Dots
indicate the position of $\zeta_{\bf k}=\mu + 2t(\cos(k_x)+\cos(k_y))$ for
the respective momentum.}
\end{figure}

\begin{figure}
\includegraphics[width=\columnwidth]{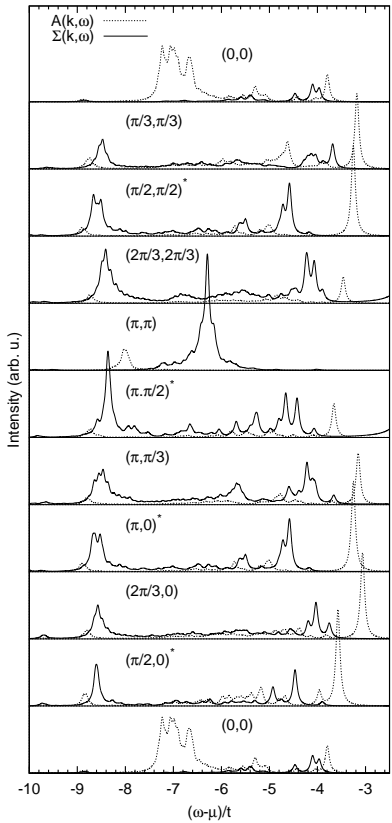}
\caption{\label{fig3}  
Spectral density $A({\bf k},\omega)$ and imaginary part of
the self-energy $\Sigma({\bf k},\omega)$ 
at half-filling. The figure combines spectra
from the $16$- and $18$-site clusters, momenta from the
$16$-site cluster are marked by asterisks. The spectra are computed
with $\eta=0.05$, $\Sigma({\bf k},\omega)$ is multiplied by $1/5$ and
at $(\pi,\pi)$ by an extra factor of $0.25$.}
\end{figure}

Figures \ref{fig2} and \ref{fig3} show
the single particle spectral function $A({\bf k},\omega)$
and  the imaginary part 
$\frac{1}{\pi}\;\Im \;\Sigma({\bf k},\omega-i\eta)$
at half-filling. Particle-hole symmetry
fixes the chemical potential at $\mu=U/2$.
Figure  \ref{fig2} shows the entire energy range of the lower and upper
Hubbard band whereas Fig.  \ref{fig3} shows a closeup of the
lower Hubbard band. Both Figures combine spectra from the $16$
and $18$ site cluster, which produces several 
${\bf k}$-points along each high-symmetry line.\\
As expected $\Sigma({\bf k},\omega)$ shows an intense peak
within the Hubbard gap which has a quite substantial dispersion.
With the exception of the peaks at $(\pi,\pi)$ and $(0,0)$ 
the dispersion of this peak is remarkably consistent with an inverted
nearest neighbor dispersion, i.e. 
$\zeta_{\bf k}-\mu=-\epsilon_{\bf k}=2t(\cos(k_x)+\cos(k_y))$ 
which is indicated in Fig.  \ref{fig2}.
There are two possible interpreatations for the deviating behaviour at
$(\pi,\pi)$ and $(0,0)$. It may be that the dominant pole 
has a very rapid dispersion of its energy $\zeta_{{\bf k},\nu}$
in the neighborhood of these momenta. A second possibility is that there
are two poles with a smooth dispersion $\zeta_{{\bf k},\nu}$
but rapidly varying $\sigma_{{\bf k},\nu}$.
The peak at $(\pi,\pi)$ and $\omega \approx -6t$ would then
belong to a second band of poles, which has appreciable residuum
only near $(\pi,\pi)$, whereas the residuum of the pole in the gap
would suddenly drop to zero at $(\pi,\pi)$.
A calculation at $U/t=20$ shows, however, that the dispersion of the central
pole including $(\pi,\pi)$ and $(0,0)$
is almost exactly the same, which makes the interpretation
in terms of a single central pole with
rapid dispersion near these momenta more plausible.\\
To understand the meaning of this form of the self-energy
let us consider a self energy with a single dispersive pole
\begin{equation}
\Sigma({\bf k},\omega) = \frac{U}{2} + \frac{\eta}{\omega -\frac{U}{2} + \epsilon_{\bf k}}.
\end{equation}
where the first term $U/2$ is the Hartree potential at half-filling.
This yields the quasiparticle dispersion
\begin{equation}
E_{\pm,{\bf k}}= \frac{U}{2} \pm \sqrt{\eta + \epsilon_{\bf k}^2}.
\label{disp}
\end{equation}
This is similar to spin density-wave mean-field
theory which would be obtained by setting  $\eta=m^2U^2/4$. 
An expansion of the type (\ref{fit}) gives the constant term
$\sigma_0=13.2$ for $U/t=10$ and $\sigma_0=84.1$ for $U/t=20$ so that
we would obtain $m=0.53$ for $U/t=10$ and $m=0.84$ for $U/t=20$.
On the other hand it is known from experiment that the
dispersion of the spectral weight of the quasiparticle band at half-filling
is not consistent with spin density-wave theory in that the
spectral weight drops sharply in the outer part of 
the zone\cite{Wells,Ronning}. This is due to additional
features in $\Sigma({\bf k},\omega)$.\\
The closeup of the lower Hubbard band in Fig. \ref{fig3}  reveals
additional structure in  $\Sigma({\bf k},\omega)$. For most momenta
there are two
essentially dispersionless `humps' at $-4t$ and $-8.5t$
with a broad continuum in between them. The upper peak at $-4t$
shows some oscillation. It turns out, however, that the reason for this
oscillation is that all peaks obtained in the $16$-site cluster are
shifted by $\approx -0.3t$ relative to those from the $18$-site cluster.
This shift and hence the entire oscillation may therefore be a finite-size
effect. Together with the dominant peak within the Hubbard gap
the peak - or group of peaks - at $\approx -4t$ encloses
the quasiparticle band at the top of the photoemission spectrum.
The intensity of the continuum 
is minimal at $(0,0)$ and increases towards the zone boundary.
Right at $(\pi,\pi)$, however, the continuum is more or less absent.
Also on this smaller
scale $\Sigma({\bf k},\omega)$ thus shows a very rapid ${\bf k}$-dependence
in the neighborhood of $(\pi,\pi)$. This behaviour is
seen consistently in all clusters studied and is not an artefact
of one specific cluster geometry. The `band' of poles at $\approx -4t$
is the reason for the deviation from the simple spin-density
wave form of the dispersion, Eq. (\ref{disp}). As can be seen in 
Fig. \ref{fig3}
the quasiparticle peak at $(0,0)$ is located immediately above the 
respective pole of the self-energy which has relatively small residuum.
As discussed above, this implies a small weight of the
quasiparticle peak itself. The dispersionless band  of poles 
at $\approx -4t$
thus reduces the bandwidth - because it cannot be crossed by the
quasiparticle band - and also the spectral weight near
$(0,0)$ and $(\pi,\pi)$.\\
We proceed to the hole doped case and consider the spectra for
the cluster ground state with 2 holes, corresponding to
$\delta=0.125$ in $16$ sites and $\delta=0.11$ in $18$ sites.
The dominant pole within the gap still
has a strong dispersion although the bandwidth 
is reduced as compared to half-filling
and the dispersion deviates from the simple inverted
nearest-neighbor-hopping dispersion.
\begin{figure}
\includegraphics[width=\columnwidth]{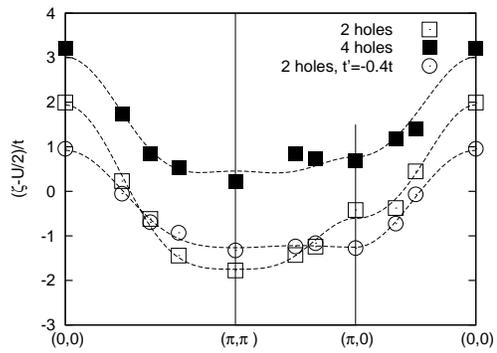}
\caption{\label{fig4} 
Dispersion of the central peak of $\Sigma({\bf k},\omega)$
in the Hubbard gap for various
doped systems. The lines are fits to tight binding harmonics with
coefficients given in Table \ref{tab1}. The value $U/t=10$ for all
systems.}
\end{figure}
\begin{table}[b]
\begin{center}
\begin{tabular}{|l|rrrr|}
\hline
 & $\zeta_{0}$ & $\zeta_{1}$ & $\zeta_{2}$ & $\zeta_{3}$ \\
\hline
2 hole, U/t=10 & 4.479 &      0.461 &      0.087 &      0.066 \\
2 hole, U/t=10$^1$ & 4.302 &      0.511 &      0.152 &      0.079 \\
2 hole, U/t=20 & 9.600 &      0.697 &      0.023 &      0.028 \\
2 hole, U/t=20$^1$ & 9.290 &      0.747 &      0.065 &      0.029 \\
2 hole, U/t=40 & 20.536 &      0.686 &     -0.022 &     -0.001 \\
4 hole, U/t=10 & 6.093 &      0.320 &      0.120 &      0.040 \\
4 hole, U/t=20 & 12.004 &      0.446 &      0.028 &      0.013 \\
2 hole, U/t=10$^2$ & 4.328 &      0.273 &      0.135 &     -0.010 \\
\hline
& $\sigma_{0}$ & $\sigma_{1}$ & $\sigma_{2}$ & $\sigma_{3}$ \\
\hline
2 hole, U/t=10 &  10.055 &      0.243 &     -0.981 &     -0.467 \\
2 hole, U/t=10$^1$ &  9.580 &      0.171 &     -1.034 &     -0.526 \\
2 hole, U/t=20 &  81.686 &      0.634 &     -0.174 &      0.045 \\
2 hole, U/t=20$^1$ &81.367 &      0.910 &     -0.235 &      0.053 \\
2 hole, U/t=40 & 376.799 &      0.396 &      0.302 &      0.396 \\
4 hole, U/t=10 &  9.625 &     -0.023 &     -0.428 &     -0.420 \\
4 hole, U/t=20 & 80.216 &     -0.312 &      0.081 &      0.094 \\
2 hole, U/t=10$^2$ &  8.489 &     -0.139 &     -0.762 &     -0.438\\
\hline
\end{tabular}
\caption{Expansion coefficients of the dispersion and residuum
of the central pole in the gap for different systems.\\
$^1$: Data from the $20$-site cluster. $^2$: Data with $t'/t=-0.4$.}
\label{tab1}
\end{center}
\end{table}
This can be seen in Fig. \ref{fig4} which shows the dispersion of 
this central peak for a few systems and in Table \ref{tab1} which gives 
the corresponding parameters $\zeta_{i}$ and $\sigma_{i}$ for a variety
of clusters. The following trends can be realized
in Table  \ref{tab1}: for two holes
the reduction of the bandwidth of the pole
saturates at approximately $0.7$ for large $U/t$. The deviations
from the simple inverted nearest-neighbor-hopping dispersion
seem to vanish in that limit. For $4$ holes the same
holds true, but the saturation value for the reduction
of the bandwidth is $0.4$. The bandwidth of the central pole
thus decreases with doping, the deviations from
the inverted nearest-neighbor hopping
dispersion vanish with increasing $U/t$.\\
The average residuum  $\sigma_{0}$ increases roughly
as $U^2$. Unlike the width of the dispersion, the
weight of the central pole seems to be rather independent
on doping. 
Figure  \ref{fig4} moreover shows that already for two
holes the peaks at $(0,0)$ and $(\pi,\pi)$ fit in smoothly
into the dispersion and this holds true for all doped systems. \\
In addition to this large peak for two holes
a second band of less intense poles of
$\Sigma({\bf k},\omega)$ appears in the neighborhood of $(\pi,\pi)$.
This can be seen in Fig. 
\ref{fig5} which shows the spectral function and the
self-energy for the lower Hubbard band.
The pole in question starts out with the intense peak at 
$\omega \approx \mu -2t$  at $(\pi,\pi)$ and then
rapidly disperses upwards. The respective peaks are
very pronounced at $(\frac{2\pi}{3},\frac{2\pi}{3} )$
and $(\pi,\frac{\pi}{3})$, somewhat less clear at
$(\pi,\frac{\pi}{2})$. 
This new band of poles can also be seen in Fig. \ref{fig6} 
which shows the corresponding spectra for the $20$-site cluster
with 2 holes. At $(\pi,\pi)$ itself the large peak is again
at approximately $\omega  = \mu -2t$ and at
the two momenta $(\frac{3\pi}{5},\frac{4\pi}{5})$
and  $(\frac{4\pi}{5},\frac{2\pi}{5})$ near $(\pi,\pi)$ the
intense peak is present as well.
Since this upward dispersing band of poles
can be identified only near $(\pi,\pi)$ the
residuum $\sigma_{{\bf k},\nu}$ of this pole must have a strong
${\bf k}$-dependence and decrease rapidly with increasing distance from
$(\pi,\pi)$.
\begin{figure}
\includegraphics[width=\columnwidth]{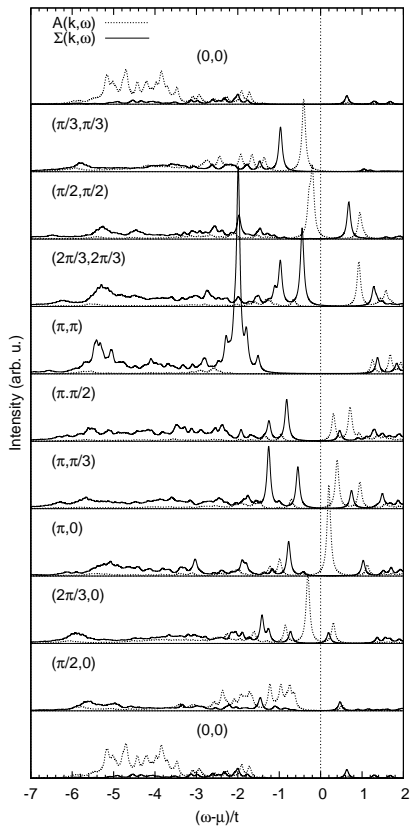}
\caption{\label{fig5} 
Spectral density $A({\bf k},\omega)$ and imaginary part of
the self-energy $\Sigma({\bf k},\omega)$ 
for the $16$-site and $18$-site cluster with $2$ holes. 
The spectra are computed
with $\eta=0.05$, $\Sigma({\bf k},\omega)$ is multiplied by $1/5$.}
\end{figure}
At $(\pi,\pi)$ itself there is now also a broad incoherent continuum
and the large peak at $\omega \approx\mu-2t$ seems to have merged with 
this continuum.
As was the case for half-filling
the intensity of the continuum increases from the center
to the edge of the Brillouin zone. Accordingly, the
remnant of the free-electron band can still be seen at $(0,0)$ -
this is the broad hump at $\approx -4.5t$ in both, Fig. \ref{fig5}
and \ref{fig6} - but is 
damped out for all other momenta.
There is actually one difference between the $20$-site cluster 
and the $16$- and $18$-site cluster:
one might assign a third band of poles at the top of the incoherent
continuum at $\omega \approx -t$ in Fig. \ref{fig5} - this band
is completely absent in the spectra for $20$-site cluster in Fig.
\ref{fig6}.
Figure \ref{fig7} shows the spectral function and self-energy for
the $20$-site cluster with two holes and $U/t=20$. This is qualitatively
the same as in Fig. \ref{fig6} but the bandwidth of
the upward dispersing peak is reduced by approximately a factor
of $2$. The dispersion of this peak thus obviously scales with
$J=4t^2/U$ and this is confirmed by other systems.\\
The presence of an upward-dispersing band of poles of
$\Sigma({\bf k},\omega)$ near $(\pi,\pi)$ and $\mu$
would be of crucial importance for the Fermi surface topology.
Since the quasiparticle band cannot cross a band of
isolated poles of $\Sigma({\bf k},\omega)$ this would
force the quasiparticle band to bend downward\cite{Stanescu,Imada1,Imada2} 
and thereby cut off the
low energy inverse photoemission weight at momenta around $(\pi,\pi)$
in the energy range $0.5t \rightarrow 2t$ in Fig. \ref{fig5}.
This is a very plausible interpretation, because 
- as  shown in Ref. \cite{inverse} - this low energy
inverse photoemission weight is not part of
a quasiparticle band, but spin-polaron shake-off.
The downward bending
of the quasiparticle band in turn would lead to a band maximum
and hence a hole-pocket-like
Fermi surface - 
as found by exact diagonalization of the 
t-J model\cite{poc1,poc2,poc3} and various version of Cluster Dynamical Mean
Field Theory\cite{Stanescu,Imada1,Imada2}.
As will be seen in a moment, this upward dispersing band of poles
in the self-energy is indeed a special feature of the underdoped
regime.\\
\begin{figure}
\includegraphics[width=\columnwidth]{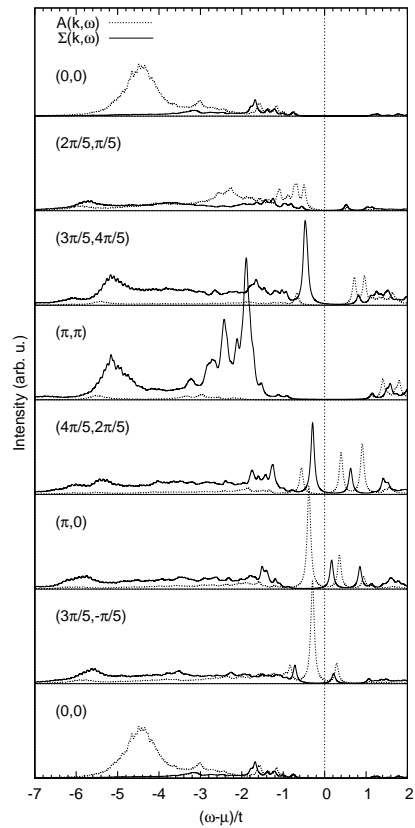}
\caption{\label{fig6} 
Spectral density $A({\bf k},\omega)$ and imaginary part of
the self-energy $\Sigma({\bf k},\omega)$ 
for the $20$-site cluster with $2$ holes. 
The spectra are computed
with $\eta=0.05$, $\Sigma({\bf k},\omega)$ is multiplied by $1/5$.}
\end{figure}
\begin{figure}
\includegraphics[width=\columnwidth]{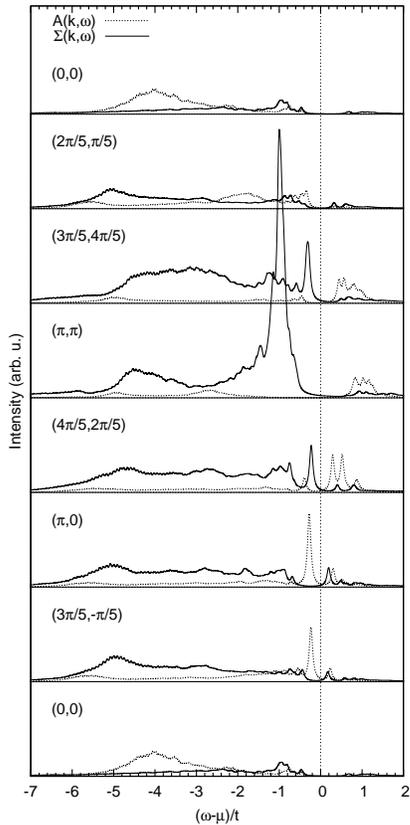}
\caption{\label{fig7} 
Same as Fig. \ref{fig6} but with $U/t=20$.}
\end{figure}
Figure \ref{fig8} shows the spectral function
and self-energy for $4$ holes, corresponding to $\delta=0.25$ in $16$ sites
and $\delta=0.22$ in $18$ sites and reveals a profound change
in the self-energy. More precisely, the upward dispersing
band of poles around $(\pi,\pi)$ which was present in the underdoped case
\begin{figure}
\includegraphics[width=\columnwidth]{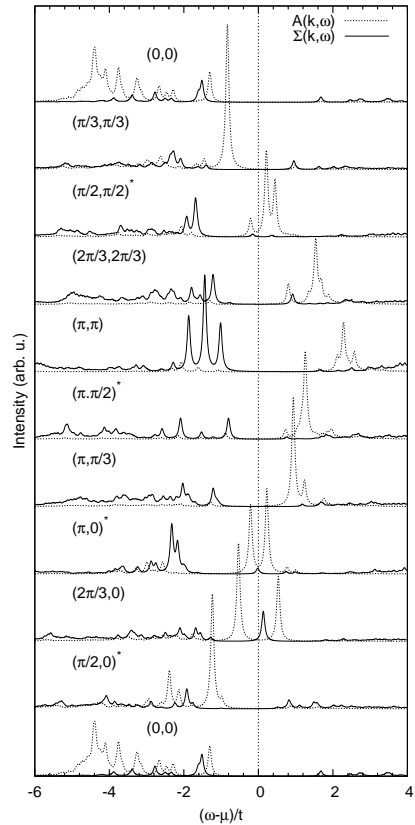}
\caption{\label{fig8} 
Spectral density $A({\bf k},\omega)$ and imaginary part of
the self-energy $\Sigma({\bf k},\omega)$ 
for the $16$-site and $18$-site
cluster with $4$ holes. The spectra are calculated with $\eta=0.05$ and
$\Sigma({\bf k},\omega)$ is multiplied by $1/5$.} 
\end{figure}
now has disappeared. The small peaks close to $\mu$
which can be seen at $(\frac{\pi}{2},\frac{\pi}{2})$, $(\pi,0)$
and $(\frac{2\pi}{3},0)$
are probably a kind of finite size effect:
at these momenta the quasiparticle peak is split between
photoemission and inverse photoemission and since
there is always a finite-size gap
between phtoemission and inverse photoemission spectrum
in a finite system this results in a two-peak structure in
the Green's function. This two-peak structure 
of the Green's function in turn necessitates
a pole of the self-energy in between. In an infinite system,
however, there is no splitting of a peak at $\mu$ by a finite amount,
so that the respective peak in $\Sigma({\bf k},\omega)$ would be absent.
Without these small peaks, however,  $\Sigma({\bf k},\omega)$ 
has no significant peak above $\approx \mu-0.8t$.
There are in fact some stronger peaks at the top of the continuum,
particularly so at $(\pi,\pi)$ and also at $(\frac{2\pi}{3},\frac{2\pi}{3})$
and $(\frac{\pi}{2},\frac{\pi}{2})$ but if one wanted to assign a
band this would rather have a shallow maximum at $(\pi,\pi)$ and then
disperse downwards as one moves towards either $(0,0)$ or $(\pi,0)$.
The upward dispersing band of poles in $\Sigma({\bf k},\omega)$
around $(\pi,\pi)$ seen in the underdoped
clusters thus is definitely absent, which
implies a `connected' nearest neighbor hopping band
which starts out at $\omega \approx \mu-t$ at $(0,0)$ and
reaches $\omega \approx \mu+2t$ at $(\pi,\pi)$. This band
will produce a  large Fermi surface but with a
band mass that is enhanced by a factor of $\approx 2.7$.\\
The further development with doping then is not really interesting
any more: the central peak persists until hole dopings of $50$\%
and becomes increasingly dispersionless, the lower Hubbard band
stays similar to Fig. \ref{fig8}.
\section{Extrapolation to the infinite system}
Our next objective is to extrapolate the cluster results to the infinite
system. It has to be noted beforehand that the results may not
be expected to be quantitatively correct - this will become apparent 
by varios numerical checks - but rather give a qualitative picture.
This is simply a consequence of the inavoidable limitations
due to the small cluster size.\\
We represent the self-energy by the following ansatz
\begin{eqnarray}
\Sigma({\bf k},\omega)&=& nU + 
\sum_{\nu=1}^3\frac{\sigma_{{\bf k},\nu}}{\omega-\zeta_{{\bf k},\nu}}
\nonumber \\
&&\;\;\;
+ \sum_{\lambda=1}^3 \sigma_{c,\lambda}({\bf k}) 
\log\left(\frac{\omega-e_{min,\lambda}}{\omega-e_{max,\lambda}}\right).
\label{ansatz}
\end{eqnarray}
The first term is the Hartree-Fock potential, the second term
describes - for the underdoped case - the three dominant poles:
the `central pole' in the gap ($\nu=1$), the upward dispersing
pole near ${\bf Q}=(\pi,\pi)$ ($\nu=2$) and the pole at the top of
the continuum ($\nu=3$). The last term is the contribution
from the incoherent continua 
which we model by a constant spectral density between  ${\bf k}$-independent
limits $e_{min}$ and $e_{max}$ but with a ${\bf k}$-dependent intensity
$\sigma_{c}({\bf k})$.
There are two such continua in the lower Hubbard band,
one below $\mu$ and the other above $\mu$, and a
third one for the upper Hubbard band.
Since pole number 2, the upward dispersing
pole near ${\bf Q}$, can be seen only for
a few momenta we terminate the expansion (\ref{fit}) after
the second term for this pole, i.e. $\zeta_{2,2}$ and  $\zeta_{3,2}$
are taken to be zero from the beginning.
This pole moreover shows a strong variation of its residuum,  
$\sigma_{{\bf k},2}$, which rapidly decreases with the distance from
$(\pi,\pi)$ so that the pole cannot be identified anymore for
more distant momenta. Accordingly we approximate  $\sigma_{{\bf k},2}$ 
as
\begin{eqnarray} 
\sigma_{{\bf k},2} &=& \sigma_{0,2} \;e^{-f({\bf k})/\alpha}
\nonumber \\
f({\bf k})&=&4\left(\cos^2\left(\frac{k_x}{2}\right) + \cos^2\left(\frac{k_y}{2}\right)\right)
\label{model_1}
\end{eqnarray}
Finally, the amplitude of the incoherent continua is written as
\begin{equation}
\sigma_c({\bf k})= \sigma_{c,0} +
\sigma_{c,1}\left(1 \pm\frac{\gamma_{1}({\bf k})}{4} \right).
\end{equation}
where the $-$ ($+$) sign refers to continua
in the lower (upper) Hubbard band.
For the overdoped regime we use the same ansatz (\ref{ansatz}) but
without the upward dispersing pole. The pole
at the top of the continuum ($\nu$=2) has a rapid variation of its residuum
as well so we use the expression (\ref{model_1}).
The coefficients which describe the dispersion of the central pole
$\nu=1$ are given in Table \ref{tab1}, the remaining
coeffcients are listed in Table \ref{tab2}.
\begin{figure}
\includegraphics[width=\columnwidth]{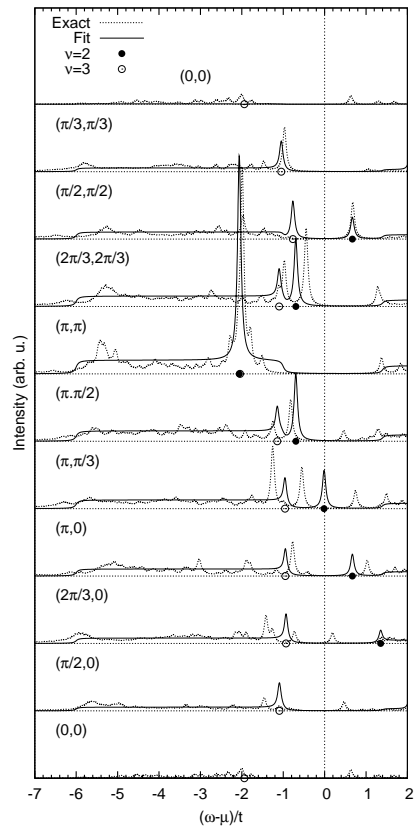}
\caption{\label{fig9} 
$\Sigma({\bf k},\omega)$ for $U/t=10$ and two holes in the $16$ and $18$-site
cluster compared to the
fit (\ref{ansatz}). The spectra are computed with $\eta=0.05$ and
$\Sigma({\bf k},\omega)$ is multiplied by $1/5$.
The dots indicate the dispersion of the poles $\nu=2$ and $\nu=3$.}
\end{figure}
\begin{table}[b]
\begin{center}
\begin{tabular}{|c|cccc|ccc|}
\hline
& $\zeta_{0,\nu}$ & $\zeta_{1,\nu}$ & $\zeta_{2,\nu}$ & $\zeta_{3,\nu}$ &
$\sigma_{0,\nu}$ & $\alpha$ & \\
\hline
$\nu=2$ & 0.670 &  0.683 &       &        & 3.347 & 1.777 &  \\ 
\hline
& $\zeta_{0,\nu}$ & $\zeta_{1,\nu}$ & $\zeta_{2,\nu}$ & $\zeta_{3,\nu}$ &
$\sigma_{0,\nu}$ & $\sigma_{2,\nu}$ 
& $\sigma_{3,\nu}$ \\
$\nu=3$ &-1.118  & 0.013 & -0.131& -0.088 & 0.407  & -0.054 & -0.051    \\
\hline
& $\epsilon_{min}$ & $\epsilon_{max}$  & $\sigma_{c,0}$ & $\sigma_{c,1}$ & & &\\
\hline
& -6.0 & -1.0 & 0.035 & 0.7 & & &\\
&  1.4 &  3.0 & 0.020 & 0.4 & & &\\
&  7.0 & 12.0 & 0.050 & 1.0 & & &\\
\hline
\hline
& $\zeta_{0,\nu}$ & $\zeta_{1,\nu}$ & $\zeta_{2,\nu}$ & $\zeta_{3,\nu}$ &
$\sigma_{0,\nu}$ & $\alpha$ & \\
\hline
$\nu=2$ & -1.617 & -0.101 & 0.130 & -0.028 & 2.500 & 2.000 &  \\
\hline
& $\epsilon_{min}$ & $\epsilon_{max}$  & $\sigma_{c,0}$ & $\sigma_{c,1}$ & & &\\
\hline
& -6.0 & -1.5 & 0.36  & 0.32 & & &\\
&  1.0 &  6.0 & 0.30  & 0.0  & & &\\
\hline
\end{tabular}
\caption{Coefficients of the model self-energy for $\delta \approx 12\%$
(top) and  $\delta \approx 24\%$ (bottom).
The coefficients for the pole $\nu=1$ are given in Table \ref{tab1}.
The constant terms $\zeta_{0,\nu}$ and the edges of the continua
are relative to $\mu$.}
\label{tab2}
\end{center}
\end{table}
Figure \ref{fig9} compares the fitted self-energy in the
underdoped case with the cluster spectrum, Fig. \ref{fig10}
shows the same comparison for the overdoped vase.
The agreement is not perfect but the fitted self-energy
reproduces the essential features. The
assignment of `bands' in the self-energy clearly
involves some degree of arbitrariness.
It should also be noted that the dispersion of the pole $\nu=2$ 
has no significance in those regions of ${\bf k}$-space where
its residuum is small. Those parts which have large residuum, however,
appear to be fitted roughly correct. The fact that the
band $\nu=2$ crosses the chemical potential leads to
additional inaccuracies: in a small cluster there is always
an artificial finite gap between the photoemission and inverse
photoemission spectrum, because the respective electron densities differ
by a finite amount. This artificial gap can be up to
$0.5t$ in the clusters studied and 
necessarily affects the dispersion of any band which crosses
$\mu$. This is certainly one reason
for the inaccuracies of the fit for the band $\nu=2$.
It therefore
has to be kept in mind that the fitted self energies may
not be expected to be quantitatively correct - rather
the purpose of the fit is to illlustrate the consequences of the
form of the self-energy in a more qualitative fashion.\\
\begin{figure}
\includegraphics[width=\columnwidth]{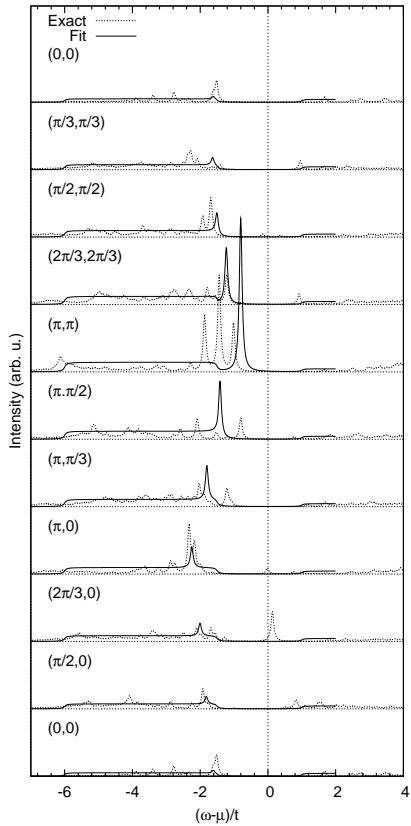}
\caption{\label{fig10} 
Self energy for $U/t=10$ and four  holes in the $16$ and $18$-site
cluster compared to the fit. }
\end{figure}
Next, we use the model self-energies to obtain approximate
single-particle-spectra for the
infinite system. For the underdoped (overdoped) regime we choose the
electron density per spin to be $n=(1-\delta)/2$
with $\delta=0.12$ ($\delta=0.24$). We fix the chemical
potential $\mu$ by demanding that the integrated spectral weight up to
$\mu$ to be equal to $n$:
\begin{equation}
\frac{1}{N}\;\sum_{\bf k}\;\int_{-\infty}^\mu A({\bf k},\omega) d\omega
=\frac{1-\delta}{2}
\label{speccon}
\end{equation}
It turns out that the $\mu$ values obtained in this
way deviate only slightly (deviation $\le 0.2t$) from
the chemical potentials of the cluster spectra.
Figure \ref{fig11}  then shows the single-particle spectral density
for $\delta=0.12$. The upper Hubbard band has been
omitted because we represented the self-energy in this energy range
only by a continuum and did not attempt to fit any fine structure.
\begin{figure}
\includegraphics[width=\columnwidth]{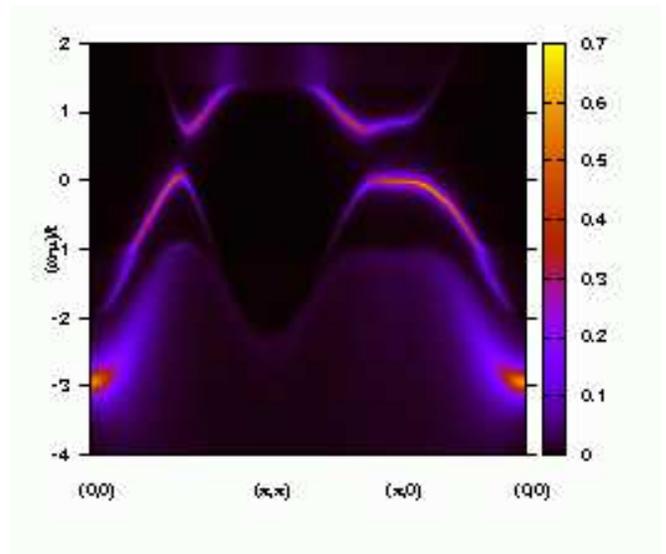}
\caption{\label{fig11} 
Single particle spectral density $A({\bf k},\omega)$
computed with the fitted self-energy (\ref{ansatz})
for the underdoped case $\delta=0.12$, the spectra are calculated
with $\eta=0.05$.}
\end{figure}
We note first that the spectral density in Fig. \ref{fig11}
is in very good agreement with the spectral density obtained by
Quantum Monte-Carlo (QMC) simulations of the underdoped 
Hubbard model, see e.g. Fig. 9 of Ref. \cite{Carsten}.
The two-band structure of the valence band, the flat high-intensity
part around $(\pi,0)$ and the apparent nearest neighbor hopping
band in the energy range $[-t:t]$ are completely consistent
with QMC. The intensity of the upper band in the photoemission
spectrum is low at $(0,0)$ and increases as the Fermi energy is approached,
whereas the lower band at $-3t$ has a high intensity at $(0,0)$
and rapidly looses 
weight as is moves away from this momentum. This is in agreement
with the QMC spectra as well. \\
\begin{figure}
\includegraphics[width=\columnwidth]{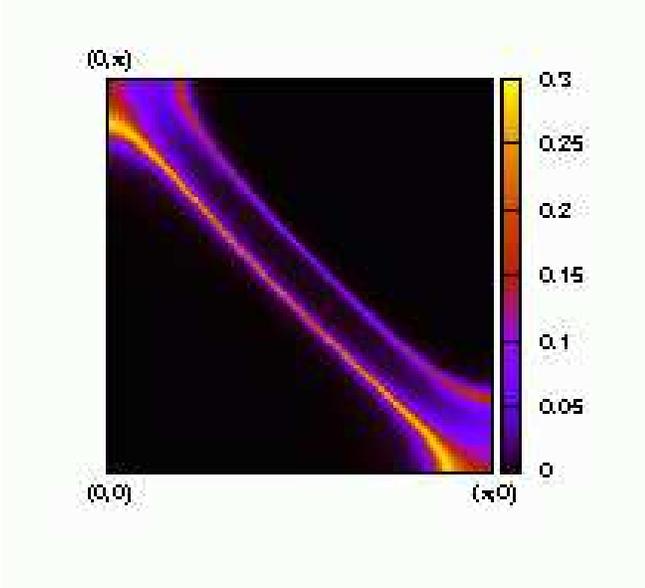}
\caption{\label{fig12} 
Single particle spectral density at $\mu$ computed with the fitted 
self-energy (\ref{ansatz}) for the underdoped case, $\delta=0.12$.
The value of $\eta$ is $0.05$.}
\end{figure}
Figure \ref{fig12} shows a Brillouin zone map
of the spectral weight at $\mu$.
The Fermi surface obviously takes the form of a `hole ring' along
the surface of the antiferromagnetic Brillouin zone.
Since the free dispersion $\epsilon_{\bf k}$ is degenerate
along the line $(\pi,0)\rightarrow(0,\pi)$ there is no hole pocket 
but a `hole ring'. Calculations for a single hole
in the t-J model usually give - for moderate
$J/t$ - a very small dispersion
along this line and a shallow maximum at 
$(\frac{\pi}{2},\frac{\pi}{2})$\cite{Trugman,Shraiman,Inoue,Ederbecker}.
If this band is filled with holes this results in hole pockets
centered at $(\frac{\pi}{2},\frac{\pi}{2})$.
The reason for the maximum
is hole hopping along a spiral path as first discussed by
Trugman\cite{Trugman}. The present calculation either
misses this fine detail or it is not relevant in the doped system 
so that no maximum exists and the pockets are deformed into
a ring.
In any way, the Fermi surface clearly is `small' in that it covers
only a tiny fraction of the Brillouin zone. This is
ultimately the consequence of the upward dispersing pole
of $\Sigma({\bf k},\omega)$ near $(\pi,\pi)$. 
The fraction of the Brillouin
zone covered by the ring is $17.3\%$. This is much larger
than the value of $\delta/2=6\%$. The latter value would be
obtained if the doped holes
are modelled by spin-$1/2$ Fermions as suggested by exact 
diagonalization\cite{lan} and as predicted in a recent theory for
lightly Mott insulators\cite{ewo}. It is quite obvious, however, that
small changes in the parameters characterizing the fitted self energy
may change this value strongly.
The too large area of the ring thus probably simply shows the
limited accuracy of the fit.
One notable feature is the
small spectral weight of the Fermi surface facing $(\pi,\pi)$ -
this is very similar to the `remnant Fermi surface'.\\
Next, we consider the overdoped case and set $n=0.38$.
Since there are no bands of poles close  to $\mu$ and in particular
the band of poles near $(\pi,\pi)$ is absent we expect
a free-electron-like Fermi surface. This is indeed the case as can be seen
from the Fermi surface map in Fig. \ref{fig13}.
The fraction of the Brillouin zone covered by the large electron-like
Fermi surface around $(0,0)$ is $0.382$, which is in very good
agreement with the Luttinger theorem when the carriers are
electrons. Taken together the data thus indicate a phase transition
in between underdoping and overdoping from a phase with
hole pockets - or a `hole ring' in the present case - to one with a 
large Fermi surface. 
\begin{figure}
\includegraphics[width=\columnwidth]{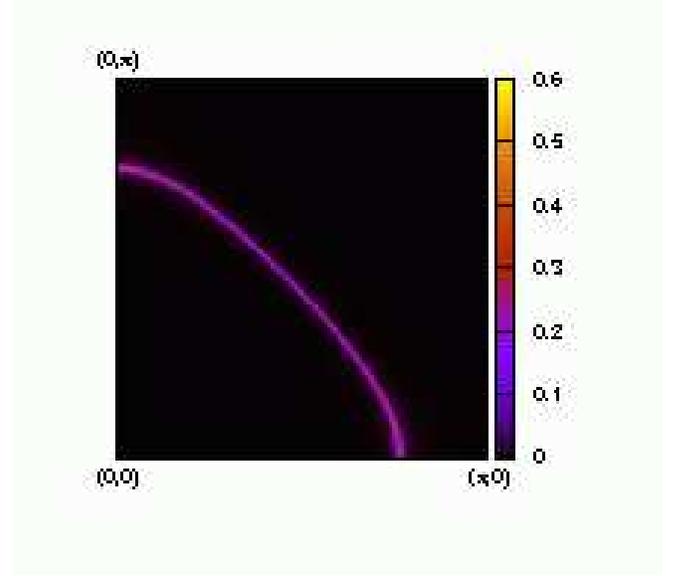}
\caption{\label{fig13} 
Single particle spectral density at $\mu$ computed with the fitted self-energy
for the overdoped case.}
\end{figure}
\section{The self-energy in the underdoped regime}
\begin{figure}
\includegraphics[width=\columnwidth]{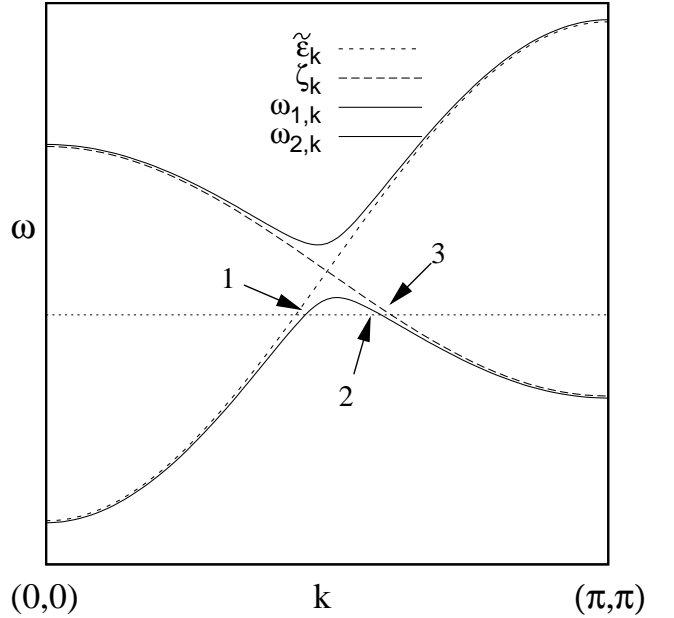}
\caption{\label{fig14} 
Bare dispersion $\tilde{\epsilon}_{\bf k}$ intersecting with a band of
poles of the self-energy, $\zeta_{\bf k}$.}
\end{figure}

We have seen that a key feature of the underdoped system is
the presence of an additional upward dispersing band of poles
in $\Sigma({\bf k},\omega)$ near $(\pi,\pi)$,
i.e. the band $\nu=2$. In the following we discuss
some consequences of this band. We set
\begin{equation}
\epsilon_{\bf k} + \Sigma({\bf k},\omega+\mu) = \tilde{\epsilon}_{\bf k }
+ \frac{\sigma}{\omega - \zeta_{\bf k }}
\end{equation}
where the first term on the r.h.s. is the sum of $\epsilon_{\bf k}$,
$g_{\bf k}$ and the contribution from the other poles in
$\Sigma({\bf k},\omega)$ and the second term represents the
upward dispersing band $\nu=2$. 
We assume that $\tilde{\epsilon}_{\bf k }$
is a smooth function of ${\bf k}$ and for simplicity neglect
its frequency dependence. In the absence of the isolated
pole $\tilde{\epsilon}_{\bf k }$ therefore
would be the quasiparticle dispersion.
If additional poles of $\Sigma({\bf k},\omega)$ are sufficiently far
away we obtain the two bands
\begin{equation}
\omega_{1,2}= \frac{1}{2}\left(
\tilde{\epsilon}_{\bf k } +\zeta_{\bf k } \pm
\sqrt{(\tilde{\epsilon}_{\bf k } -\zeta_{\bf k })^2+ 4\sigma}\right)
\end{equation}
which are shown in Fig. \ref{fig14}.
There is a gap of $2\sqrt{\sigma}$ between these bands.
Far from the crossing point $\tilde{\epsilon}_{\bf k } =\zeta_{\bf k }$
the bands take the form
\begin{eqnarray}
\omega_{1,2}&=& \tilde{\epsilon}_{\bf k } + \frac{\sigma}{\tilde{\epsilon}_{\bf
    k } -\zeta_{\bf k }} \nonumber \\
\omega_{1,2}&=&  \zeta_{\bf k } - \frac{\sigma}{\tilde{\epsilon}_{\bf
    k } -\zeta_{\bf k }} 
\end{eqnarray}
The two resulting
bands thus  partially trace the quasiparticle band $\tilde{\epsilon}_{\bf k }$
and partially the dispersion of the pole, $\zeta_{\bf k }$.
The spectral weight of the respective branches is
\begin{eqnarray}
Z^{-1}&=& 1 - \frac{\partial \tilde{\epsilon}_{\bf k }}{\partial \omega}
+ \frac{\sigma}{(\tilde{\epsilon}_{\bf k } -\zeta_{\bf k })^2}
\nonumber \\
Z^{-1}&=& 1 - \frac{\partial \tilde{\epsilon}_{\bf k }}{\partial \omega}
+ \frac{(\tilde{\epsilon}_{\bf k } -\zeta_{\bf k })^2}{\sigma}
\end{eqnarray}
where the upper (lower) line refers to the band portion tracing
$\tilde{\epsilon}_{\bf k }$ ($\zeta_{\bf k }$).
Since far from the crossing point
$(\tilde{\epsilon}_{\bf k } -\zeta_{\bf k })^2\gg \sigma$
the spectral weight $Z$ assumes its usual value 
for the band portion tracing $\tilde{\epsilon}_{\bf k }$
but is $\frac{\sigma}{(\tilde{\epsilon}_{\bf k } -\zeta_{\bf k })^2}\ll 1$
for the band portion tracing $\zeta_{\bf k }$.
This behaviour can be seen along $(0,0)\rightarrow (\pi,\pi)$
and along $(\pi,0)\rightarrow (\pi,\pi)$ in Fig. \ref{fig11}.
The upward dispersing band of poles in $\Sigma({\bf k},\omega)$
also has a major significance for the
validity of the Luttinger theorem\cite{Luttingertheorem}. 
To see this we derive a slightly modified version of the theorem
which allows for an appealing physical interpretation. We consider
\begin{eqnarray}
S&=&\frac{1}{2\pi i}\sum_{\bf k} \;\int_C \;d\omega\;
G({\bf k},\omega)\;
\frac{\partial \Sigma({\bf k},\omega)}{\partial \omega} 
\nonumber \\
&=&\frac{1}{2\pi i}\sum_{\bf k} \;\int_C \;d\omega\;\left(\;
G({\bf k},\omega) + \frac{1}{G({\bf k},\omega)}\;
\frac{\partial G({\bf k},\omega)}{\partial \omega}\;\right)
\nonumber \\
\label{sdef}
\end{eqnarray}
where $C$ is a curve in the complex $\omega$-plane which encloses
the part of the real axis with $\omega < \mu$
in counterclockwise fashion.
All singularities of the integrand are located on the real axis.
The first term in the second line will give the 
total electron number/spin. The second term has two kinds of singularties:
poles and zeros of $ G({\bf k},\omega)$.
Near a pole we have
\begin{eqnarray}
\frac{1}{G({\bf k},\omega)}\;
\frac{\partial G({\bf k},\omega)}{\partial \omega}
&\approx& \frac{ -\frac{Z_{{\bf k},\nu}}{(\omega-\omega_{{\bf k},\nu})^2}}
{\frac{Z_{{\bf k},\nu}}{(\omega-\omega_{{\bf k},\nu})}} + \dots
\nonumber \\
= -\frac{1}{(\omega-\omega_{{\bf k},\nu})} + \dots
\end{eqnarray}
whereas near a zero we have
\begin{eqnarray}
\frac{1}{G({\bf k},\omega)}\;
\frac{\partial G({\bf k},\omega)}{\partial \omega}
&\approx& \frac{ \sigma_{{\bf k},j} }
{\sigma_{{\bf k},j}(\omega-\zeta_{{\bf k},j})} + \dots
\nonumber \\
= \frac{1}{(\omega-\zeta_{{\bf k},j})} + \dots
\end{eqnarray}
It follows that
\begin{eqnarray}
N_e&=& 2S + 2 \sum_{\bf k} m_{\bf k}\nonumber \\
m_{\bf k} &=& \sum_\nu \Theta(\mu -\omega_{{\bf k},\nu})  - 
\sum_j \Theta(\mu -\zeta_{{\bf k},j}).
\label{rawversion}
\end{eqnarray}
If we assume that $S=0$ we find that the number of electrons
can be obtained by computing the number of
`occupied' poles of the Green's function and subtracting the number
of `occupied' poles of the self-energy.
A band of poles of the self-energy which crosses $\mu$ - such as the
band $\nu=2$ introduced in the fit of the self-energy in the preceeding
section - therefore
produces a `negative volume Fermi surface' because the number 
of momenta within this surface has to be subtracted in the computation 
of the electron number.\\
To arrive at the known form of the theorem we note that
- as discussed in section II - there is exactly
one pole of $G({\bf k},\omega)$
between any two successive poles of  $\Sigma({\bf k},\omega)$. Moreover
it is easy to see that there is always precisely one pole of 
$G({\bf k},\omega)$ below
the lowest pole of $\Sigma({\bf k},\omega)$. Accordingly
$m_{\bf k}=0$ if the topmost singularity below $\mu$ is a pole of the 
self-energy
for the respective ${\bf k}$-point and $m_{\bf k}=1$
if the topmost singularity is a pole of the Green's function.
Since in the first case $\Re\; G({\bf k},\mu)<0$ 
whereas in the second case
$\Re\;G({\bf k},\mu)>0$ we obtain 
\begin{equation}
N_e = 2\sum_{\bf k} \Theta( \Re\;G({\bf k},\mu) ) 
\label{endversion}
\end{equation}
which is the `generalized' Luttinger theorem given by
Dzyaloshinskii\cite{Dzyalo}. The equivalence of (\ref{rawversion})
and (\ref{endversion}) has previously been noted by 
Ortloff {\em et al.}\cite{Ortloff}.\\
It is then easy to see that - contrary to widespread belief -
hole pockets can in fact be completely consistent
with the Luttinger theorem. We again consider
Fig. \ref{fig14} which shows a situation where
the dispersion $\tilde{\epsilon}_{\bf k}$ is intersected by an
upward dispersing band of poles
of the self-energy, $\zeta_{\bf k}$, resulting in the two
quasiparticle bands $\omega_{1,{\bf k}}$ and $\omega_{2,{\bf k}}$.
The lower of these bands, $\omega_{1,{\bf k}}$, 
crosses the Fermi energy - indicated by the horizontal dashed
line - and produces two Fermi level
crossings at the points $1$ and $2$ which form the hole pocket.
In between $(0,0)\rightarrow 1$ we have $m_{\bf k}=1$ because
the topmost pole below $\mu$ is the pole $\omega_{1,{\bf k}}$ of
the Greens function. Along $1\rightarrow 2$ $m_{\bf k}=0$ because
no pole of either Green's function nor self-energy
is below $\mu$.  Additional singularities at lower energies
do not change this: since the topmost singularity below
$\omega_{1,{\bf k}}$ must be a pole of the self-energy, the
total contribution to $m_{\bf k}$ from all singularities
including this one is zero. Along the short piece
$2\rightarrow 3$ we have again $m_{\bf k}=1$
but along $3\rightarrow (\pi,\pi)$ we have $m_{\bf k}=0$
because the topmost pole below $\mu$ now is one of the self-energy,
$\zeta_{\bf k}$. The piece $2\rightarrow 3$ will be very short if the
residuum $\sigma$ is small. The piece $3\rightarrow (\pi,\pi)$
corresponds precisely to the `negative volume Fermi surface' 
discussed above because
here the topmost singularity is a pole of the self-energy. Assuming that
the total volume of the hole pockets is $V_{BZ}\;\delta/2$ - as suggested
by a recent theory of the lightly doped Mott insulator\cite{ewo} - the
fraction of the Brillouin zone outside the pockets is
$V_{BZ}\;(1-\delta/2)$. Then, if the  `negative volume Fermi surface'
is $V_{BZ}/2$ the occupied part of the Brillouin zone in the sense of the
Luttinger theorem would be $V_{BZ}(1-\delta)/2)$,
i.e. corresponding precisely to the electron density. 
Hole pockets therefore would be
completely consistent with the Luttinger theorem if this is
applied correctly. For example, evaluation of
(\ref{endversion}) with the fitted self-energy for two holes -
where the Fermi surface takes the form of a hole ring, 
see Fig. \ref{fig12} -
gives $N_e=1.032N$ whereas the correct value would be
$N_e=0.88N$. The deviation of $\approx 15\%$
shows the limited accuracy of the
fitted self-energy, but is far smaller than the difference
in Fermi surface volume.\\
\section{Comparison to ARPES experiments}
ARPES experiments on underdoped cuprate superconductors show a number
of interesting features and the next point is a comparison of the
extrapolated cluster spectra to these experiments.
Here an important point is to introduce longer range hopping terms.
More precisely, we introduce an additional hopping term which connects 
next-nearest
(i.e. $(1,1)$-like) neighbors. We choose the matrix element for
this term to be $t'=-0.4t$. This value is somewhat large but we
simultaneously
omit a hopping term between $(2,0)$-like neighbors because this
would lead to a strong increase in the number of three-site
terms in the strong coupling Hamiltonian.
We performed the calculation only for the
$16$- and $18$-site 
cluster with two holes because introduction of this additional
hopping term increases the number of possible three-site combinations
in the Hamiltonian considerably, so that the calculations become
too difficult for the $20$-site cluster with two holes
and the $18$-site cluster with $4$ holes. It turns out that the ground 
states of the two
clusters with the $t'$-term in the Hamiltonian
have somewhat unusual quantum numbers: the ground state of
two holes in the $16$-site cluster has momentum $(\pi,0)$ and
spin $S=0$, the ground state of the $18$-site cluster with two holes
has momentum $(\frac{2\pi}{3},0)$
and spin $S=1$. This means that the GS of the $16$-site cluster is twofold,
that of the $18$-site cluster $12$-fold degenerate. This presents no 
real problem in that the expression (\ref{green}) should be
viewed as the zero temperature limit of a grand canonical average, so that
in the case of GS-degeneracy one simply has to average over all
degenerate ground states.\\
\begin{figure}
\includegraphics[width=\columnwidth]{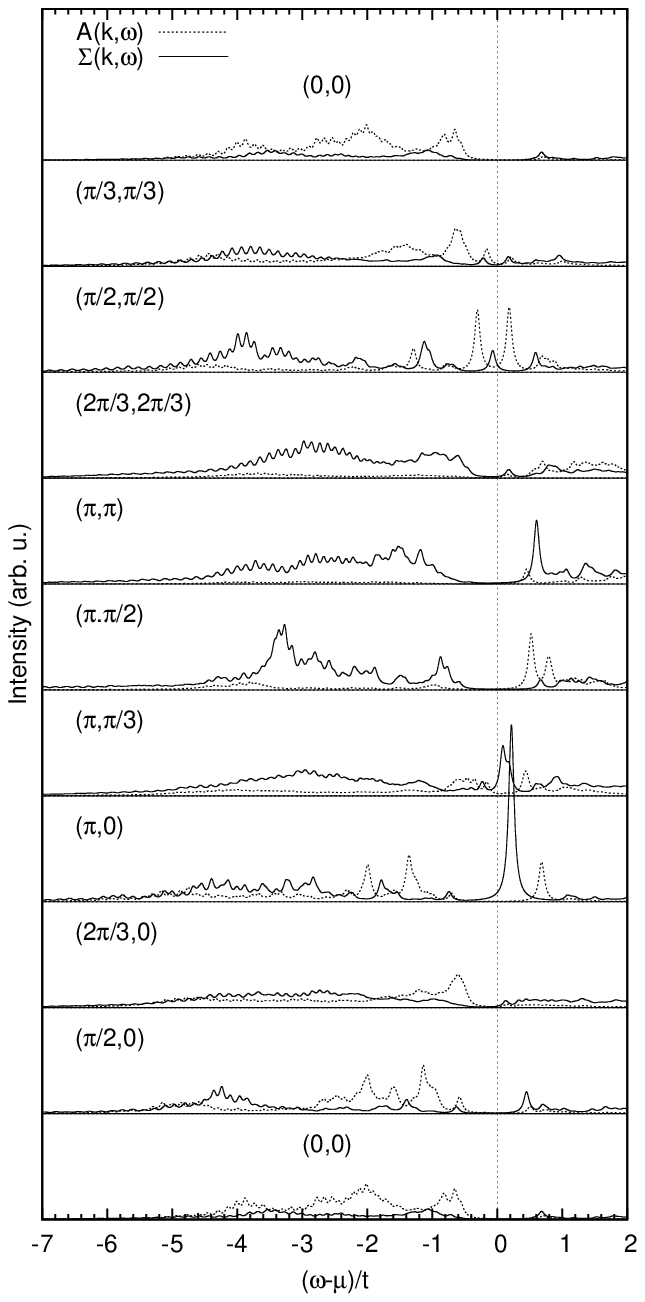}
\caption{\label{fig15} 
Spectral density $A({\bf k},\omega)$ and imaginary part of
the self-energy $\Sigma({\bf k},\omega)$ 
for the $16$-site and $18$-site
cluster with $2$ holes. The value $t'=-0.4 t$. The spectra are calculated
with $\eta=0.05$ and $\Sigma({\bf k},\omega)$ is multiplied by 1/5.} 
\end{figure}
From Fig. \ref{fig4} and  Table \ref{tab1}
it can be seen that the dispersion of the 
central pole within the gap $(\nu=1)$ is changed somewhat by
the presence of $t'$. In particular
the deviations from the inverted nearest-neighbor-hopping dispersion
become stronger.
Figure \ref{fig15} shows the single particle spectrum and self-energy
in the presence of the $t'$-term. By comparison with Fig. \ref{fig5}
it is obvious that the $t'$-term introduces several pronounced
changes in $\Sigma({\bf k},\omega)$: The residuum of the
pole near $(\pi,\pi)$ ($\nu=2$) has decreased considerably,
in fact this pole cannot seen anymore but takes the form of
broad humps at the top of the incoherent continua, 
whose upper edge has moved
closer to $\mu$. 
Instead there are now poles with large residuum
at $(\pi,0)$ and $(\pi,\frac{\pi}{3})$. And finally there is no more pole
at the top of the incoherent continuum and the intensity
of the incoherent continua themselves has increased. To fit the self-energy
we use the same ansatz (\ref{ansatz}) but drop the pole $\nu=3$
at the top of the incoherent continuum.
The second difference concerns the
strong pole near $(\pi,0)$. We assume that this
pole actually belongs to the band $\nu=2$ and to model this we change
the residuum of this band of poles by adding a second term:
\begin{eqnarray}
\sigma_{{\bf k},2} &=& \sigma_0 \;e^{-f({\bf k})/\alpha}
+  \sigma_d\; e^{\gamma_d^2/\beta^2}\;\gamma_d^2({\bf k}),
\nonumber \\
\gamma_d({\bf k})&=& \cos(k_x)-\cos(k_y).
\label{tppole}
\end{eqnarray}
The dispersion of this pole
is again expanded with respect to only two harmonics,
the constant $\gamma_{0}({\bf k})$ and the nearest-neighbor-hopping
harmonic $\gamma_{1}({\bf k})$. As already mentioned the
poles near $(\pi,\pi)$ cannot really be resolved in the calculated
self-energies. Our main justification for keeping this band is
the behaviour seen for $t'=0$.\\
\begin{table}[b]
\begin{center}
\begin{tabular}{|c|cccccc|}
\hline
& $\zeta_{0,\nu}$ & $\zeta_{1,\nu}$ & $\sigma_{0}$ & $\alpha$& $\sigma_{d}$ & $\beta$\\
\hline
$\nu=2$ & 0.143  &  0.325  & 0.400  & 6.8  & 0.0184 & 1.15  \\ 
\hline
& $\epsilon_{min}$ & $\epsilon_{max}$  & $\sigma_{c,0}$ & $\sigma_{c,1}$ & & \\
\hline
& -5.0 & -0.5 & 0.25 & 0.5 & & \\
&  0.5 &  3.0 & 0.020 & 0.4 & & \\
&  7.0 & 12.0 & 0.025 & 1.0 & & \\
\hline
\end{tabular}
\caption{Coefficients of the model self-energy for 
$U/t=10$, $\delta \approx 10\%$, $t'/t=-0.4$. 
The constant term $\zeta_{0,2}$ and the edges of the continua
are relative to $\mu$.
A comparison of the model self-energy
and the numerical spectrum can be seen in Fig. \ref{fig13}. }
\label{tab4}
\end{center}
\end{table}
Table \ref{tab4} then gives the respective coefficients and
Fig. \ref{fig13} compares the numerical self-energy and the fit.
The coefficient of $\gamma_{1}({\bf k})$ is positive.
If only the term $\propto \sigma_d$ were kept in (\ref{tppole})
the exponential $e^{\gamma_d^2/\alpha_2^2}$ replaced by unity
and $\zeta_{0,\nu}$ be set to zero
in this term,  the resulting self-energy would be 
identical to the phenomenological self-energy introduced by
Yang {\em et al.}\cite{YangRiceZhang}. 
On the other hand  the residuum clearly has an extremely strong
${\bf k}$-dependence so that the exponential cannot be
neglected in the present fit. \\
\begin{figure}
\includegraphics[width=\columnwidth]{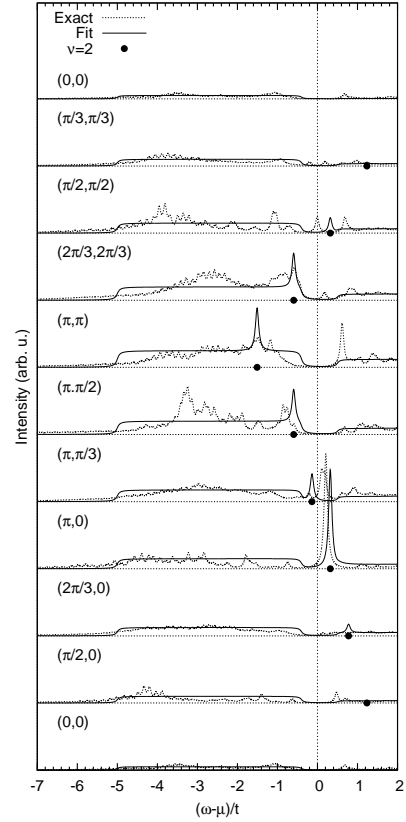}
\caption{\label{fig16} 
Imaginary part of
the self-energy $\Sigma({\bf k},\omega)$ 
for the $16$-site and $18$-site
cluster with $2$ holes and Fit. The value $t'=-0.4 t$.} 
\end{figure}
Figure \ref{fig17} then shows the single particle spectral density obtained
with the fitted self-energy.
Along $(0,0)\rightarrow (\pi,\pi)$ the quasiparticle band disperses 
towards the Fermi energy.
Consistent with experiment, the intensity of the band thereby increases
as $\mu$ is approached.
\begin{figure}
\includegraphics[width=\columnwidth]{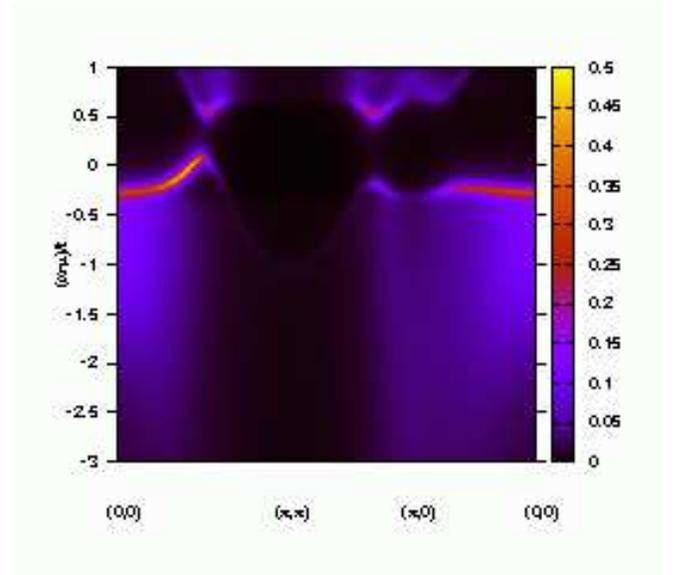}
\caption{\label{fig17} 
Single particle spectral density $A({\bf k},\omega)$
for $\delta=0.12$, $U/t=10$ and $t'/t=-0.4$ obtained by using the
interpolated self-energy for the infinite lattice.}
\end{figure}
After crossing $\mu$ the band turns downward sharply
and immediately crosses $\mu$ again whereby
its spectral weight drops. This is precisely the situation shown in
Fig. \ref{fig14}.
Along $(0,0)\rightarrow (\pi,0)$ the band disperses upward 
as well but does not reach $\mu$. 
The ARPES spectrum thus shows a
`pseudogap' but this is a trivial consequence of the
Fermi surface being a hole pocket centered at 
$(\frac{\pi}{2},\frac{\pi}{2})$ (see below).
Roughly at $(\frac{4\pi}{5},0)$ there is a maximum of the dispersion
with high spectral weight and the band turns downward and looses
weight beyond this point. This behaviour may actually have been observed 
by Chuang {\em et al.}\cite{Chuang} who interpreted this as
indicating a Fermi level crossing at $\approx(\frac{4\pi}{5},0)$.
Chuang {\em et al.} observed this behaviour 
in an underdoped compound - see Fig. 2i of
Ref. \cite{Chuang} - but also in overdoped samples where it is
unclear if it can be compared to the present calculation.
In contrast the spectral weight around $(\pi,0)$ is small. This is
consistent with experiment where a quasiparticle band around $(\pi,0)$
is usually not observed in the normal state of underdoped
cuprate superconductors. Along the
line $(\pi,0)\rightarrow (\pi,\pi)$ the band seems to disperse
upward at first, but then again bends sharply and disperses away from $\mu$.
At the turning point the spectral weight drops. This is again
due to the avoided crossing of the quasiparticle band
with the upward dispersing band of poles of $\Sigma({\bf k},\omega)$ 
near $(\pi,\pi)$, i.e. the band $\nu=2$, in Fig. \ref{fig14}. 
A very similar behaviour has in fact been observed experimentally
in underdoped La$_{2-x}$Sr$_x$CuO$_4$, see Fig. 5 of Ref. \cite{Ino}.
In this compound the quasiparticle band is sufficiently far from
$\mu$ at $(\pi,0)$ so that the absence of a Fermi level crossing
is obvious for $x=0.05$ and $x=0.10$. A very avoided crossing
has been observed as `backbending' of bands in Bi2201 
in Ref. \cite{Hashimoto}.
If such an avoided crossing would occur
sufficiently close to $\mu$, however, it may look very similar to a 
true Fermi level crossing. Aparent experimental Fermi level
crossings along $(\pi,0)\rightarrow (\pi,\pi)$ thus should be
considered with care.\\
The actual Fermi surface of the underdoped system is
shown in Fig. \ref{fig18} and takes the form of a hole
pocket, centered near $(\frac{\pi}{2},\frac{\pi}{2})$.
The pocket
is shifted slightly towards $(0,0)$ and the part facing
$(\pi,\pi)$ has smaller spectral weight and less curvature than the part
facing $(0,0)$. The pockets covers $1.86\%$ of the total Brillouin zone
which would correspond - assuming twofold spin degeneracy and four
equaivalent pockets - to a quasiparticle
density of $14.9\%$. This is close to the hole concentration
of $12\%$. On the other hand the electron density
as computed from the Luttinger theorem (\ref{endversion})
is $1.25$ so that the inaccuracies of the self energy clearly are
substabtial and the close agreement for the quasiparticle density may be
fortuitious.
\begin{figure}
\includegraphics[width=\columnwidth]{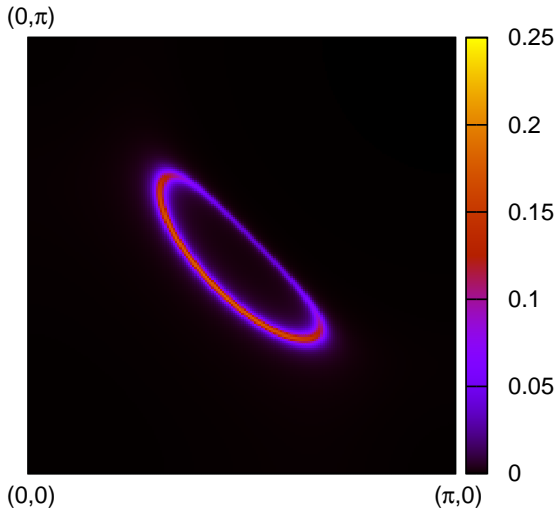}
\caption{\label{fig18} 
Single particle spectral density at $\mu$
for $\delta=0.12$, $U/t=10$ and $t'/t=-0.4$ obtained by using the
interpolated self-energy for the infinite lattice.}
\end{figure}
\section{Discussion}
In summary we have presented an exact diagonalization study of the
self-energy $\Sigma({\bf k},\omega)$
in the 2D Hubbard model. Larger clusters than usual could be
used because instead of studying the true Hubbard model we considered
its strong-coupling limit which requires much smaller Hilbert spaces.
For dopings less than 30\% - i.e. the doping region in which
cuprate superconductivity takes place - 
several distinct features can be identified: first, a pole with
large residuum $\propto (U/t)^2$ and a 
dispersion of width $\propto t$ in the center of the Hubbard gap,
which is present throughout this doping range. Second,
a pole with smaller residuum and an upward dispersion $\propto J$
around $(\pi,\pi)$ which is present only in the underdoped regime.
And third, several broad incoherent continua of width
$\propto t$. The top of the lowest of these incoherent
continua may be formed by a third pole. 
All features in the spectral representation of
$\Sigma({\bf k},\omega)$ show
a pronounced ${\bf k}$-dependence, both with respect to their
dispersion and their residuum. This implies that in real space
the self-energy is long-ranged and oscillatory. \\
The key difference between the underdoped and overdoped
system is the presence of a dispersive pole of
$\Sigma({\bf k},\omega)$ around $(\pi,\pi)$. This pole cuts
through the quasiparticle dispersion and changes the
Fermi surface topology completely.  In the underdoped hole concentration 
range the Fermi surface takes the form of a `hole ring' for
$t'=0$ or hole pockets centered near $(\frac{\pi}{2},\frac{\pi}{2})$
for $t'=-0.4t$ and changes
to a large free-electron-like Fermi surface in the overdoped case. 
We have shown that the hole pockets can be completely consistent with
the Luttinger theorem.
The residuum of the upward dispersing pole
near $(\pi,\pi)$  thus plays the role of an order parameter
for the phase transition between the two different ground states.\\
The single-particle spectra obtained with the fitted self-energies
agree very well with both Quantum Monte Carlo
simulations (for $t'=0$) and ARPES on underdoped cuprates
(for $t'=0.4 t$). In particular the spectra reproduce a hole-pocket-like
Fermi surface of similar shape and location as in experiment
and also the characteristic strong
asymmetry of the spectral weight of the parts of the pocket 
facing $(0,0)$ and $(\pi,\pi)$.
Quite generally, the spectra show that all parts of the quasiparticle
band which deviate
from the noninteracting electron band structure have a very small
weight. The reason is that these band portions closely
follow the dispersion of a
band of poles of  $\Sigma({\bf k},\omega)$ and `borrow' their spectral weight
from the quasiparticle band.
The data suggest in particular that some
Fermi level crossings observed in ARPES along the line
$(\pi,0)\rightarrow (\pi,\pi)$ actually may not be true Fermi
level crossings but sharp bends at the intersection of
the quasiparticle dispersion and a band of poles
in $\Sigma({\bf k},\omega)$. An example where such pseudo-crossings
can be clearly recognized is La$_{2-x}$Sr$_x$CuO$_4$\cite{Ino}.
In this compound the shift the quasiparticle band is sufficiently
far from $\mu$ near $(\pi,0)$ so that the 
absence of a Fermi level crossing is clear - if 
the quasiparticle band is closer to $\mu$, however, there
pseudo-crossing may well be mistaken for a real Fermi level crossing.\\
An interesting question is the physical meaning
of the upwards dispersing pole
near $(\pi,\pi)$. One immediate consequence of this pole is that the
part of the inverse photoemission spectrum belonging to
the lower Hubbard band consists of two disconneted
components. The first component is
the unoccupied part of the quasiparticle band, which forms
the cap of the hole-pockets around $(\frac{\pi}{2},\frac{\pi}{2})$,
the second component is
a disconnected part around $(\pi,\pi)$. This two-component nature
of the inverse phtoemission spectrum was discussed
in Ref. \cite{inverse} where it was shown that the disconnected
component around $(\pi,\pi)$ actually consists of spin-polaron shakeoff, i.e.
spin excitations which are released when a hole dressed by
antiferromagnetic spin fluctuations is filled by an electron.
The disconnected nature of the low energy inverse photoemission
spectrum thus is a quite natural consequence of the hole pockets
and the spin-polaron nature of the quasiparticles.\\
The transition then may be understood as follows:
in a Mott insulator the electrons are localized and retain only their
spin degrees of freedom. 
Upon doping most electrons are still tightly surrounded by other electrons
as in the insulator and thus remain localized - the ground state
wave function therefore should optimize the energy gain due to 
delocalization 
of the holes and since the holes are spin-1/2 Fermions
this can be achieved by forming hole pockets with a volume $\propto \delta$
around the ground state momentum of a single hole i.e. 
$(\frac{\pi}{2},\frac{\pi}{2})$. This picture leads to a simple
theory\cite{ewo} in which the number of noninteracting particles is equal to
the  number of doped holes. \\
When the density of holes becomes sufficiently large so that the
electrons are sufficiently mobile
a phase transition occurs to a state which optimizes
the kinetic energy of the electrons themselves and this means
the formation of a Fermi surface with a volume $\propto 1-\delta$.
A rough estimate for the hole concentration where the transition occurs would
be $\delta_c \approx z^{-1}$ with $z$ the
coordination number because then
each electron has one unoccupied neighbor on the average.
The vanishing of the upward dispersing pole in the self-energy then
corresponds to this phase transition.
Experimentally it seems that the phase transition between pockets
and free-electron-like Fermi surface occurs right at optimal doping.
This suggests that superconductivity is also related to this
transition.\\
Acknowledgement: K. S. acknowledges the JSPS Re-
search Fellowships for Young Scientists. R. E. most
gratefully acknowledges the kind hospitality at the Cen-
ter for Frontier Science, Chiba University. This work
was supported in part by a Grant-in-Aid for Scientific
Research (Grant No. 22540363) From the Ministry of
Education, Culture, Sports, Science and Technology of
Japan. A part of the computations was carried out at the
Research Center for Computational Science, Okazaki Re-
search Facilities and the Institute for Solid State Physics,
University of Tokyo.

\section{Appendix A}
Here we proove that the real constant $g_{\bf k}$ in (\ref{self})
is equal to the Hartree-Fock potential. Considering the
limit of large $\omega$, expanding the two expressions for
the Green's function, (\ref{self}) and (\ref{green}), in powers of
$\omega^{-1}$ thereby using (\ref{spectral}) one obtains by comparing
the terms $\propto \omega^{-2}$:
\begin{equation}
\epsilon_{\bf k} + g_{\bf k} =
\langle \Psi_0| \left[ c_{{\bf k},\sigma}^{} [H, c_{{\bf k},\sigma}^\dagger] 
\right]_+|\Psi_0\rangle
\end{equation}
where $[\dots]_+$ denotes the anticommutator.
Using  a standard Hamiltonian of the form
\begin{eqnarray*}
H&=&\sum_{{\bf k},\sigma} \epsilon_{\bf k} c_{{\bf k},\sigma}^\dagger
c_{{\bf k},\sigma}^{}\nonumber \\
&+& \frac{1}{2} \sum_{{\bf k},{\bf k}',{\bf q}}\sum_{\sigma,\sigma'}
V_{{\bf k}+{\bf q},{\bf k},{\bf k}'-{\bf q},{\bf k}'}^{\sigma\sigma'}
c_{{\bf k}+{\bf q},\sigma}^\dagger c_{{\bf k}'-{\bf q},\sigma'}^\dagger 
c_{{\bf k}',\sigma'}^{}c_{{\bf k},\sigma}^{}
\end{eqnarray*}
this gives
\begin{equation}
g_{\bf k} = \sum_{{\bf k}',\sigma'}
V_{{\bf k},{\bf k}',{\bf k}',{\bf k}}^{\sigma,\sigma'} \langle n_{{\bf k}',\sigma'} \rangle -
\sum_{{\bf k}'}
V_{{\bf k},{\bf k}',{\bf k},{\bf k}'}^{\sigma,\sigma} 
\langle n_{{\bf k}',\sigma} \rangle.
\end{equation}
where $\langle \dots \rangle$ denotes the ground state expectation value.
\section{Appendix B}
The canonical transformation which reduces the full
Hubbard Hamiltonian to the strong-coupling Hamiltonian
takes the form
\begin{equation}
O'=e^{S}O e^{-S}
\label{trans}
\end{equation}
where $O$ denotes an operator in the original Hilbert space
and $O'$ is the transformed operator.
The antihermitean generator $S$ is
\begin{eqnarray}
S &=& -\sum_{i,j}\sum_\sigma\; \frac{t_{ij}}{U} \;
\left( \hat{d}_{i,\sigma}^\dagger \hat{c}_{j,\sigma}^{} -
\hat{c}_{i,\sigma}^\dagger \hat{d}_{i,\sigma}^{}\right)\nonumber \\
\hat{d}_{i,\sigma}^\dagger &=& c_{i,\sigma}^\dagger n_{i,\bar{\sigma}}\nonumber \\
\hat{c}_{i,\sigma}^\dagger &=& c_{i,\sigma}^\dagger (1-n_{i,\bar{\sigma}}).
\end{eqnarray}
The strong-coupling Hamitonian or corrected t-J model
is obtained by transforming the Hubbard Hamiltonian according to (\ref{trans})
and discarding terms of 2$^{nd}$ or higher order in $1/U$. The complete -
and somewhat lengthy - Hamiltonian in given in equation (14) of
the paper by Eskes {\em et al.}\cite{Eskesetal}.
The transformed version of the electron annihilation operator is
\begin{eqnarray}
e^{S}  c_{i,\uparrow}^{} e^{-S} &=& c_{i,\uparrow}^{} - \sum_j \frac{t_{ij}}{U} 
(\;\hat{d}_{j,\downarrow}^\dagger - \hat{c}_{j,\downarrow}^\dagger)\;
c_{i,\downarrow}^{}c_{i,\uparrow}^{} \nonumber \\
&+&
( \hat{c}_{j,\downarrow}^{} - \hat{d}_{j,\downarrow}^{})\;S_i^-
+ n_{i,\downarrow}(2n_{i,\uparrow}-1)\hat{c}_{j,\uparrow}^{}
\nonumber \\
&-& (1-n_{i,\downarrow})(2n_{i,\uparrow}-1)\hat{d}_{j,\uparrow}^{}\;)
\label{optr}
\end{eqnarray}
where again terms of higher order in $t/U$ have been neglected.
By collecting the terms
which give a nonvanishing result when acting on a state without
double occupancies and do not produce a double occupancy themselves
we obtain the operator for photoemission in the lower Hubbard 
band\cite{Eskesetal}:
\begin{equation}
\tilde{c}_{i,\uparrow}^{} = \hat{c}_{i,\uparrow}^{} -\sum_j \;\frac{t_{ij}}{U}\;
\left(\hat{c}_{j,\uparrow}^{}n_{i,\downarrow}
- \hat{c}_{j,\downarrow}^{} S_i^-\right).
\label{pesop}
\end{equation}
Since the transformed creation operator is the Hermitean conjugate of
(\ref{optr}), the operator for inverse photoemission with final
states in the lower Hubbard band is just the Hermitean conjugate of 
(\ref{pesop}).
The operator for inverse photoemission with final states in the upper
Hubbard band is obtained by taking the Hermitean conjugate of (\ref{optr})
and collecting terms which create a double occupancy but do
not annihilate one:
\begin{eqnarray}
\tilde{c}_{i,\uparrow}^\dagger &=& \hat{d}_{i,\uparrow}^{}  
+ \sum_j \;\frac{t_{ij}}{U}\;
(\;\;\hat{c}_{j,\downarrow}^{} \;c_{i,\downarrow}^\dagger \;c_{i,\uparrow}^\dagger  
\nonumber \\
&+&  \hat{d}_{j,\downarrow}^\dagger \;S_i^+ +
\hat{d}_{j,\uparrow}^\dagger\;(1-n_{i, \downarrow})(2n_{i,\uparrow}-1)\;).
\end{eqnarray}


\begin{thebibliography}{}
\bibitem{Hussey}
 N. E. Hussey, M. Abdel-Jawad, A. Carrington, A. P. Mackenzie, 
and L. Balicas, 
Nature {\bf 425}, 814 (2003).
\bibitem{Plate}
M. Plate, J. D. F. Mottershead, I. S. Elfimov, D. C. Peets, R. Liang, 
D. A. Bonn, W. N. Hardy, S. Chiuzbaian, M. Falub, M. Shi, L. Patthey, 
and A. Damascelli, Phys. Rev. Lett. {\bf 95}, 077001 (2005).
\bibitem{Vignolle}
B. Vignolle, A. Carrington, R. A. Cooper, M. M. J. French, 
A. P. Mackenzie, C. Jaudet, D. Vignolles, Cyril Proust, N. E. Hussey
Nature {\bf 455}, 952 (2008).
\bibitem{Damascelli}
A. Damascelli, Z. Hussain, and Z.-X. Shen, 
Rev. Mod. Phys. {\bf 75}, 473 (2003).
\bibitem{Wells}
B. O. Wells, Z.-X. Shen, A. Matsuura, D. M. King, M. A. Kastner, M. Greven, 
and R. J. Birgeneau, 
Phys. Rev. Lett.  {\bf 74}, 964 (1995).
\bibitem{Ronning}
F. Ronning, C. Kim, D. L. Feng, D. S. Marshall, A. G. Loeser, L. L. Miller, 
J. N. Eckstein, L. Bozovic, and Z.-X. Shen, Science {\bf 282}, 2067 (1998).
\bibitem{Uchida}
S. Uchida, T. Ido, H. Takagi, T. Arima, Y. Tokura, and S. Tajima
Phys. Rev. B {\bf 43}, 7942 (1991).
\bibitem{Padilla}
W. J. Padilla, Y. S. Lee, M. Dumm, G. Blumberg, S. Ono, K. Segawa, 
S. Komiya, Y. Ando, and D. N. Basov
Phys. Rev. B {\bf 72}, 060511 (2005). 
\bibitem{Ong}
N. P. Ong, Z. Z. Wang, J. Clayhold,
J. M. Tarascon, L. H. Greene, and W. R. McKinnon 
Phys. Rev. B {\bf 35}, 8807 (1987). 
\bibitem{Takagi}
H. Takagi, T. Ido, S. Ishibashi, M. Uota, S. Uchida, and Y. Tokura,
Phys. Rev. B {\bf 40}, 2254 (1989).
\bibitem{Ando}
Y. Ando, Y. Kurita, S. Komiya, S. Ono, and K. Segawa,
Phys. Rev. Lett.  {\bf 92}, 197001 (2004).
\bibitem{Doiron}
N. Doiron-Leyraud, C. Proust, D. LeBoeuf, J. Levallois,
J.-B. Bonnemaison, R. Liang, D.A. Bonn, W.N. Hardy, and L. Taillefer, 
Nature {\bf 447}, 565  (2007).
\bibitem{Sebastian_1}
S. E. Sebastian, N. Harrison, E. Palm, T. P. Murphy, C. H. Mielke, R. Liang, 
D. A. Bonn, W. N. Hardy, and G. G. Lonzarich,
 Nature {\bf 454}, 200 (2008).
\bibitem{Jaudet}
C. Jaudet, D. Vignolles, A. Audouard, J. Levallois, D. LeBoeuf, 
N. Doiron-Leyraud, B. Vignolle, M. Nardone, A. Zitouni, R. Liang, D. A. Bonn, 
W. N. Hardy, L. Taillefer, and C. Proust, 
Phys. Rev. Lett. {\bf 100}, 187005 (2008).
\bibitem{Audouard}
A. Audouard, C. Jaudet, D. Vignolles, R. Liang, D. A. Bonn, W. N. Hardy, 
L. Taillefer, and C. Proust,
Phys. Rev. Lett. {\bf 103}, 157003 (2009).
\bibitem{Yelland}
E. A. Yelland , J. Singleton , C. H. Mielke , N. Harrison, F. F. Balakirev, 
B. Dabrowski , J. R. Cooper,
Phys. Rev. Lett. {\bf 100}, 047003  (2008).
\bibitem{Bangura}
A. F. Bangura, J. D. Fletcher, A. Carrington, J. Levallois, M. Nardone, 
B. Vignolle, P. J. Heard, N. Doiron-Leyraud, D. LeBoeuf, L. Taillefer, 
S. Adachi, C. Proust, and N. E. Hussey,
 Phys. Rev. Lett. {\bf 100}, 047004 (2008).
\bibitem{LeBoeuf}
D. LeBoeuf, N. Doiron-Leyraud, J. Levallois, R. Daou, J.-B. Bonnemaison, 
N. E. Hussey, L. Balicas, B. J. Ramshaw, R. Liang, D. A. Bonn, W. N. Hardy, 
S. Adachi, C. Proust, and L. Taillefer,
Nature {\bf 450}, 533 (2007).
\bibitem{Chang}
J. Chang, R. Daou, Cyril Proust, David LeBoeuf, Nicolas Doiron-Leyraud, 
Francis Laliberte, B. Pingault, B. J. Ramshaw, Ruixing Liang, D. A. Bonn, 
W. N. Hardy, H. Takagi, A. B. Antunes, I. Sheikin, K. Behnia, and 
Louis Taillefer,
Phys. Rev. Lett. {\bf 104}, 057005 (2010).
\bibitem{Hinkov}
V. Hinkov, B. Keimer, A. Ivanov, P. Bourges, Y. Sidis, and
C. D. Frost, arXiv:1006.3278v1
\bibitem{Lawler}
M. J. Lawler, K. Fujita, Jhinhwan Lee, A.R. Schmidt, Y. Kohsaka, 
Chung Koo Kim, H. Eisaki, S. Uchida, J.C. Davis, J.P. Sethna, and Eun-Ah Kim
Nature {\bf 466}, 347, (2010). 
\bibitem{poc1}
R. Eder and Y. Ohta, Phys. Rev. B {\bf 51}, 6041 (1995).
\bibitem{poc2}
P. W. Leung, Phys. Rev. B {\bf 65}, 205101 (2002).
\bibitem{poc3}
P. W. Leung, Phys. Rev. B {\bf 73}, 014502 (2006).
\bibitem{r1}
E. Dagotto and J. R. Schrieffer, Phys. Rev. B {\bf 43}, 8705 (1991).
\bibitem{r2}
R. Eder, Y. Ohta, and T. Shimozato, Phys. Rev. B {\bf 50}, 3350 (1994).
\bibitem{r3}
R. Eder and Y. Ohta, Phys. Rev. B {\bf 50}, 10043 (1994).
\bibitem{lan}
S. Nishimoto, Y. Ohta, and R. Eder
Phys. Rev. B {\bf 57}, R5590 (1998).
\bibitem{den1}
T. Tohyama, P. Horsch, and S. Maekawa, Phys. Rev. Lett. {\bf 74}, 980 (1995).
\bibitem{den}
R. Eder, Y. Ohta, and S. Maekawa, Phys. Rev. Lett. {\bf 74}, 5124 (1995).
\bibitem{intermediate}
R. Eder and Y. Ohta, Phys. Rev. B {\bf 51}, 11683 (1995).
\bibitem{beckervoijta}
 M. Vojta and K. W. Becker, Europhys. Lett. {\bf 38}, 607 (1997).
\bibitem{Luttinger}
J. M. Luttinger, Phys. Rev.  {\bf 121}, 942 (1961).
\bibitem{dagoreview}
E. Dagotto, Rev. Mod. Phys. {\bf 66}, 763 (1994).
\bibitem{Hubbard}
J. Hubbard, Proc. Roy. Soc. London, Ser. A 277, {\bf 237} (1964), 
and {\bf 281}, 401 (1964).
\bibitem{ortolani}
E. Dagotto, F. Ortolani, and D. Scalapino, 
Phys. Rev. B {\bf 46}, 3183 (1992).
\bibitem{leung}
P. W. Leung, Z. Liu, E. Manousakis, M. A. Novotny, and P. E. Oppenheimer,
Phys. Rev. B {\bf 46}, 11779 (1992).
\bibitem{HarrisLange}
A.B. Harris and R.V. Lange, Phys. Rev. {\bf 157}, 295 (1967).
\bibitem{Chao}
K. A. Chao, J. Spalek and A. M. Oles, J. Phys. C 10, {\bf L271} (1977).
\bibitem{MacDo}
A. H. MacDonald, S. M. Girvin and D. Yoshioka, 
Phys. Rev. B {\bf 37}, 9753 (1988).
\bibitem{Eskesetal}
H. Eskes, A. M. Oles, M. B. J. Meinders, and W. Stephan,
Phys. Rev. B {\bf 50}, 17980 (1994).
\bibitem{EskesOles}
H. Eskes and A. M. Oles, Phys. Rev. Lett. {\bf 73}, 1279 (1994).
\bibitem{EskesEder}
H. Eskes and R. Eder, Phys. Rev. B {\bf 54}, R14226 (1996).
\bibitem{Stanescu}
T. D. Stanescu and G. Kotliar, 
Phys. Rev. B  {\bf 74}, 125110 (2006).
\bibitem{Imada1}
S. Sakai, Y. Motome, and M. Imada, 
Phys. Rev. Lett. {\bf 102}, 056404, (2009).
\bibitem{Imada2}
S. Sakai, Y. Motome, and M. Imada, 
Phys. Rev. B {\bf 82}, 134505 (2010).
\bibitem{inverse}
R. Eder and Y. Ohta, Phys. Rev. B {\bf 54}, 3576 (1996).
\bibitem{Carsten}
C. Gr\"ober, R. Eder, and W. Hanke, Phys. Rev. B {\bf 62}, 4336 (2000).
\bibitem{Trugman}
S. A. Trugman, Phys. Rev. B {\bf 37}, 1597 (1988).
\bibitem{Shraiman}
B. I. Shraiman and E. D. Siggia, 
Phys. Rev. Lett. {\bf 60}, 740 (1988).
\bibitem{Inoue}
J. Inoue and S. Maekawa, 
J. Phys. Soc. Jpn. {\bf 59}, 2110 (1989).
\bibitem{Ederbecker}
R. Eder and K. W. Becker, 
Z. Phys. B {\bf 78}, 219 (1990).
\bibitem{ewo}
R. Eder, P. Wrobel, and Y. Ohta, 
Phys. Rev. B {\bf 82}, 155109 (2010).  
\bibitem{Luttingertheorem}
J. M. Luttinger, Phys. Rev. {\bf xx}, 
\bibitem{Dzyalo}
I. Dzyaloshinskii, Phys. Rev. B {\bf 68}, 085113 (2003).
\bibitem{Ortloff}
J. Ortloff, M. Balzer, and M. Potthoff,
Eur. Phys. J. B {\bf 58}, 37 (2007).
\bibitem{YangRiceZhang}
K.-Y. Yang, T. M. Rice, and F.-C. Zhang, 
Phys.  Rev. B {\bf 73}, 174501 2006.
\bibitem{Chuang}
Y.-D. Chuang, A. D. Gromko, D. S. Dessau, Y. Aiura, Y. Yamaguchi, K. Oka, 
A. J. Arko, J. Joyce, H. Eisaki, S. I. Uchida, K. Nakamura, and Yoichi Ando, 
Phys. Rev. Lett. {\bf 83}, 3717, (1999).
\bibitem{Ino}
A. Ino, C. Kim, M. Nakamura, T. Yoshida, T. Mizokawa, A. Fujimori, 
Z.-X. Shen, T. Kakeshita, H. Eisaki, and S. Uchida,
Phys. Rev. B {\bf 65}, 094504 (2002). 
\bibitem{Hashimoto}
M. Hashimoto, R.-H. He, K. Tanaka, J.-P. Testaud, W. Meevasana, 
R. G. Moore, D. Lu, H. Yao, Y. Yoshida, H. Eisaki,
T. P. Deveraux, Z. Hussain, and Z.-X. Shen,
Nature Physics {\bf 6}, 414 (2010).
\end{thebibliography}
\end{document}